\documentclass[aos,preprint]{imsart}

\RequirePackage{amsthm,amsmath,amsfonts,amssymb}
\RequirePackage[numbers]{natbib}
\usepackage{mathrsfs}
\usepackage{amsfonts}
\usepackage{makeidx}         % allows index generation
\RequirePackage{graphicx}        % standard LaTeX graphics tool
\usepackage{multicol}        % used for the two-column index
\usepackage[bottom]{footmisc}% places footnotes at page bottom
\usepackage{epsfig}
\usepackage{bm}
\usepackage{enumitem}
\usepackage{epstopdf}
\usepackage{multirow}
\usepackage{booktabs}
\usepackage{mathtools}
\usepackage{wrapfig}
\usepackage[pagewise, mathlines]{lineno}
\usepackage{caption}
\usepackage{subcaption}
\usepackage{lscape}
\usepackage[colorinlistoftodos]{todonotes}
\usepackage{float}
\usepackage{titlesec}
\usepackage{verbatim}
\usepackage{rotating}

\usepackage[allcolors=blue,colorlinks=true]{hyperref}
\newtheorem{theorem}{Theorem}[section]
\newtheorem{lemma}[theorem]{Lemma}
\newtheorem{corollary}[theorem]{Corollary}
\newtheorem{condition}[theorem]{condition}
\theoremstyle{remark}

\newlength{\tempheight}
\newlength{\tempwidth}

\newcommand{\rowname}[1]% #1 = text
{\rotatebox{90}{\makebox[\tempheight][c]{\textbf{#1}}}}

\newcommand{\columnname}[1]% #1 = text
{\makebox[\tempwidth][c]{\textbf{#1}}}

\mathtoolsset{showonlyrefs}
\startlocaldefs
\theoremstyle{remark}
\newtheorem{algorithm}{Algorithm}
\newtheorem{remark}{Remark}
\endlocaldefs

\begin{document}

\begin{frontmatter}
\title{Efficiency of Observed Information Adaptive Designs}
%\title{A sample article title with some additional note\thanksref{T1}}
\runtitle{Observed Information Adaptive Designs}
%\thankstext{T1}{A sample of additional note to the title.}

\begin{aug}
%%%%%%%%%%%%%%%%%%%%%%%%%%%%%%%%%%%%%%%%%%%%%%
%%Only one address is permitted per author. %%
%%Only division, organization and e-mail is %%
%%included in the address.                  %%
%%Additional information can be included in %%
%%the Acknowledgments section if necessary. %%
%%%%%%%%%%%%%%%%%%%%%%%%%%%%%%%%%%%%%%%%%%%%%%
\author{\fnms{Adam} \snm{Lane}\ead[label=e1]{adam.lane@cchmc.org}},
%%%%%%%%%%%%%%%%%%%%%%%%%%%%%%%%%%%%%%%%%%%%%%
%% Addresses                                %%
%%%%%%%%%%%%%%%%%%%%%%%%%%%%%%%%%%%%%%%%%%%%%%
\address{University of Cincinnati}
\end{aug}

\begin{abstract}
In this work the primary objective is to maximize the precision of the maximum likelihood estimate in a linear regression model through the efficient design of the experiment. One common measure of precision is the unconditional mean square error. Unconditional mean square error has been a primary motivator for optimal designs; commonly, defined as the design that maximizes a concave function of the expected Fisher information. The inverse of expected Fisher information is asymptotically equal to the mean square error of the maximum likelihood estimate. There is a substantial amount of existing literature that argues the mean square error conditioned on an appropriate ancillary statistic better represents the precision of the maximum likelihood estimate. Despite evidence in favor of conditioning, limited effort has been made to find designs that are optimal with respect to conditional mean square error. The inverse of observed Fisher information is a higher order approximation of the conditional mean square error than the inverse of expected Fisher information \cite{Efro:Hink:Asse:1978}. In light of this, a more relevant objective is to find designs that optimize observed Fisher information. Unlike expected Fisher information, observed Fisher information depends on the observed data and cannot be used to design an experiment completely in advance of data collection. In a sequential experiment the observed Fisher information from past observations is available to inform the design of the next observation. In this work an adaptive design that incorporates observed Fisher information is proposed for linear regression models. It is shown that the proposed design is more efficient, at the limit, than any fixed design, including the optimal design, with respect to conditional mean square error. 
\end{abstract}



\end{frontmatter}

\section{Introduction} \label{sec:Intro}

Fisher showed that data reduction can be achieved by conditioning on an appropriate ancillary statistic without a loss of information \cite{Fish:TwoN:1934}. Briefly, suppose independent responses $y_{1},\ldots,y_{n}$ are observed from from distribution $f_{\eta}(y) = f_{0}(\varepsilon)$, where $y = \eta + \varepsilon$. The data can be reduced to the sufficient statistic $(\hat{\eta},\boldsymbol{a})$, where $\hat{\eta}$ is the maximum likelihood estimate (MLE) of $\eta$ and $\boldsymbol{a} = [y_{(2)} - y_{(1)},\ldots,y_{(n)} - y_{(n-1)}]^{T}$ is the \textit{ancillary configuration statistic}. A statistic is \textit{ancillary} if its distribution is independent of the model parameters. In this setting $\hat{\eta}$ is not sufficient and the distribution of the MLE does not contain all of the information available in the data. Conditioning on $\boldsymbol{a}$ reduces the sample space to a \textit{relevant subset} and the full information is recovered in the distribution $\hat{\eta}|\boldsymbol{a}$. This discussion reveals that there are aspects of the observed sample that inform the precision of the MLE. Conditioning on relevant subsets has received significant attention \cite{Cox:Some:1958,Fish:Samp:1961,Efro:Hink:Asse:1978,McCu:Cond:1992,Fras:Anci:2004,Ghos:Reid:Fras:Anci:2010}.

The preceding discussion was presented for a one-dimensional parameter $\eta$. In this work a linear regression model with a multi-dimensional parameter is considered. The extension to linear regression is discussed below. In this work it is assumed that the arguments in favor of conditioning on relevant subsets are accepted. Specifically, it is assumed, as argued in \cite{Efro:Hink:Asse:1978}, that the conditional mean square error (MSE) represents a more relevant measure of the precision of the MLE than the unconditional MSE. Accepting the arguments for conditioning implies that it is more appropriate to define the efficiency of a design in terms of the conditional MSE than the unconditional MSE. Below, it is discussed how traditional optimal designs should be considered efficient with respect to unconditional MSE; whereas the design proposed in this work is efficient with respect to conditional MSE. 

\subsection{The Model}

Throughout it is assumed that the responses are observed from the linear regression model
\begin{align} \label{eq:model}
y = \boldsymbol{\beta}^{T} f_{x}(x) + \varepsilon,
\end{align}
where $y$ is the univariate response; $\boldsymbol{\beta}$ is a $p\times 1$ vector within the parameter space $B$ and $B$ is an open subset of $\mathbb{R}^{p}$ .  

For design, it is assumed that independent replicates can be observed at any $x\in\mathcal{X}$ at the experimenters discretion; where $x$ is an $s\times 1$ vector of experimental factors; the design region, $\mathcal{X}$, is a compact subset of $\mathbb{R}^{s}$; and $f_{x}$ is a known mapping from $\mathcal{X}$ to $\mathbb{R}^{p}$. An exact design, denoted $\xi_{n} =\{(x_{i},w_{i})\}_{i=1}^{d}$, is comprised of a set of $d$ support points, $x_{i}$, with corresponding allocation weights, $w_{i} = n_{i}/n$, for $i=1,\dots,d$, with total sample size $n = \sum n_{i}$. Let $\boldsymbol{y_{i}} = (y_{i1},\ldots,y_{in_{id}})^{T}$ be the responses observed with support $x_{i}$ and $\boldsymbol{y} = (\boldsymbol{y_{1}}^{T},\ldots,\boldsymbol{y_{d}}^{T})^{T}$  be the response vector from a sample of size $n$.  It is further assumed that the error vector, $\boldsymbol{\varepsilon} = (\varepsilon_{11},\ldots,\varepsilon_{dn_{d}})^{T}$, is a sequence of independent and identically distributed random variables satisfying the location family condition that $f_{\eta}(y) = f_{0}(\varepsilon)$. The response function is denoted $\eta_{i}=\boldsymbol{\beta}^{T} f_{x}(x_{i})$ when the meaning is clear. 

The joint distribution of $\boldsymbol{y}_{i}$ can be factored as
\begin{equation} \label{eq:trans_dist}
f_{\eta}(\boldsymbol{y}_{i}) = c(\boldsymbol{a_{i}})g_{\boldsymbol{a_{i}}}(\hat{\eta}_{i} - \eta_{i}), \quad i=1,\ldots,d,
\end{equation}
where $\hat{\eta}_{i}$ is the MLE of $\eta_{i}$, $g_{\boldsymbol{a}_{i}}(\hat{\eta}_{i} - \eta_{i}) = f_{\eta}(\hat{\eta}_{i}|\boldsymbol{a_{i}})$ is the conditional distribution of $\hat{\eta}_{i}$ and $c(\boldsymbol{a_{i}})$ is a normalizing constant that ensures the right hand side of \eqref{eq:trans_dist} integrates to one with respect to $\hat{\eta}_{i}$ given $\boldsymbol{a_{i}}$ \cite{Fish:TwoN:1934}. Assuming that derivatives and integrals are exchangeable, the expected and observed Fisher information, with respect to $\eta_{i}$ are
\begin{equation} \label{eq:singleObsInfo}
\mathscr{F}_{i} = {\rm{E}}\left[ \frac{\partial^{2} \log  g_{\boldsymbol{a}_{i}}(\hat{\eta}_{i} - \eta_{i}) }{\partial\eta_{i}^{2}}  \right] \quad \mbox{ and } \quad \boldsymbol{i}_{\boldsymbol{a_{i}}} = \left[ \frac{\partial^{2} \log  g_{\boldsymbol{a_{i}}}(\hat{\eta}_{i} - \eta_{i}) }{\partial\eta_{i}^{2}} \right]_{\eta_{i} = \hat{\eta}_{i}},
\end{equation}
respectively. In the location family $\mathscr{F}_{i}$ does not depend on $\eta_{i}$ which implies that the expected Fisher information is constant across the design region; i.e. $\mathscr{F} = \mathscr{F}_{1} = \cdots = \mathscr{F}_{d}$. Further, $\boldsymbol{i}_{\boldsymbol{a}_{i}}$ is a function of the data through $\boldsymbol{a}_{i}$ alone and is ancillary.

In model \eqref{eq:model} the responses depend on the $p$-dimensional vector of model parameters, $\boldsymbol{\beta}$. In Section \ref{sec:Model} details are given to show that the observed and expected Fisher information with respect to $\boldsymbol{\beta}$ can be written as
\begin{align} \label{eq:ObsInfo}
J_{\boldsymbol{A}}(\boldsymbol{x}) = F^{T} I_{\boldsymbol{A}} F \quad \mbox{ and } \quad
\mathcal{F}(\xi_{n}) = \mathscr{F} F^{T}WF,
\end{align}
respectively, where $F = [f_{x}(x_{1})^{T},\ldots,f_{x}(x_{d})^{T})]^{T}$ is a $d \times p$ matrix, $W = diag(w_{1},\ldots,w_{d})$, $I_{\boldsymbol{A}} = diag(\boldsymbol{i}_{\boldsymbol{a_{1}}},\ldots,\boldsymbol{i}_{\boldsymbol{a_{d}}})$ and $\boldsymbol{A}=(\boldsymbol{a}_{1},\ldots,\boldsymbol{a_{d}})$ is the ancillary configuration matrix. At times it is necessary to differentiate between random variables and their observed realizations. This creates tension with the convention of using capital letters to represent random variables and lower case letters for their observed realizations. To ease this tension let $\mathcal{A}$ represent the random variable with corresponding ancillary configuration matrix $\boldsymbol{A}$. 

\subsection{Optimal Design}
The field of optimal design has primarily focused on optimizing the  normalized expected Fisher information, defined as $M(\xi_{n}) = \mathcal{F}(\xi_{n})/\mathscr{F}$. Normalized expected Fisher information is, generally, a matrix and cannot be optimized with respect to every objective. Instead, the optimization is with respect to a concave optimality criterion, denoted $\Psi$. Let $\mathcal{S}_{+}^{p}$ be the set of symmetric positive semidefinite matrices generated by model \eqref{eq:model}, i.e., $\mathcal{S}_{+}^{p}= \{M(\xi):\xi\in\Xi\}$, where $\Xi$ is the set of all possible designs. The criterion $\Psi$ is a mapping from $\mathcal{S}_{+}^{p}$ to $\mathbb{R}^{+}:=[0,\infty)$. Here, expected Fisher information, $M$, is said to be \emph{optimized} if $\Psi(M)$ is the maximum for all $M\in\mathcal{S}_{+}^{p}$. The design that corresponds to the optimized expected Fisher information is referred to as the \emph{optimal design}. Equivalently, the optimal design can be considered efficient with respect to expected Fisher information.

The optimality criterion is selected according to the primary experimental objective. For example, if it is desired to minimize the volume of the confidence ellipsoid of the MLE the $D$-optimal design criterion, $\Psi(M)=|M|^{1/p}$, is applied. An optimal design is analogous to $\xi_{n}$; it is a collection of $d$ optimal support points $\boldsymbol{x}^{*} =(x_{1}^{*},\ldots,x_{d}^{*})^{T}$ and corresponding optimal allocation weights $\boldsymbol{w}^{*}=(w_{1}^{*},\ldots,w_{d}^{*})^{T}$. Section \ref{sec:Design} describes optimal design in greater detail.

A fixed design is defined as any design determined in advance of the experiment. Throughout, the optimal design will be referred to as the \emph{fixed optimal design (FOD)} in order to emphasize that it is not adaptive. In this work it is accepted that the FOD is the most efficient fixed design. However, the FOD makes no account of the observed sample's impact on the precision of the MLE. The design proposed in this work adaptively uses the ancillary information in the observed sample to improve the precision of the MLE. Before this design is formally introduced, certain aspects of conditioning following adaptive designs need to be addressed.

\subsection{Conditioning in Linear Models}

In the one-dimensional parameter location family the inverse of observed Fisher information is a higher order approximation of the conditional MSE than the inverse of expected Fisher information \cite{Efro:Hink:Asse:1978}. In Section \ref{sec:Model} this result is extended to linear regression, under conditions given in Section \ref{sec:Tech}, as
\begin{align}
{\rm{MSE}}[\boldsymbol{\hat{\beta}}|\mathcal{A}] &= [J_{\mathcal{A}}(\boldsymbol{x})]^{-1}[1 + O_{p}(n^{-1})] \quad \mbox{and} \label{eq:orderI} \\
{\rm{MSE}}[\boldsymbol{\hat{\beta}}|\mathcal{A}] &= [\mathcal{F}(\xi_{n})]^{-1}[1 + O_{p}(n^{-1/2})] \label{eq:orderF}.
\end{align}
The above establishes that in linear models the inverse observed Fisher information is a second order approximation to the conditional MSE; whereas the inverse of expected Fisher information is only a first order approximation. First and second order are defined as having error $O_{p}(n^{-1/2})$ and $O_{p}(n^{-1})$, respectively. Accepting that the conditional MSE better represents the precision of the MLE implies that the observed Fisher information is preferred to the expected Fisher information. 

The majority of the results on conditioning on relevant subsets are in the context of post-data inference following fixed designs. There are several challenges that must be addressed for this type of conditioning to be compelling for adaptive designs. The first challenge is the stigma associated with conditional inference following adaptive designs. After the completion of an adaptive procedure the design is not always ancillary and conditioning on the design may result in a loss of information \cite{Ford:Titt:Wu:infe:1985,Rose:Lach:Rand:2002}. A second challenge is whether the arguments in favor of conditioning on relevant subsets are as compelling following an adaptive design. The third challenge is that adaptive designs are known to induce a dependence in the responses. This dependence can negatively impact inference; bias and non-approximate normality of the parameter estimates are potential side effects \cite{Lane:Flou:2012,Lane:Yao:Flou:Info:2014,Lane:Wang:Flou:Cond:2016,Tari:Flou:Dist:2019,Tari:Flou:Effe:2019,Flou:Oron:Bias:2019}. A final challenge is whether the inverse of observed Fisher information still represents a higher order approximation of the conditional MSE following an adaptive design. Each of the above points are addressed in Section \ref{sec:Cond} and it is shown that conditioning on relevant subsets is just as valid and compelling following the design proposed in Section \ref{sec:adapt} as it is following a fixed design. %Further, it confirms that following the design proposed in this work the inverse of observed Fisher information is a higher order approximation to the conditional MSE than the inverse of the expected information. 

\subsection{Adaptive Designs}

The inverse of expected Fisher information is a second order approximation of the unconditional MSE \cite{McCu:Tens:1987}. From the conditional perspective the inverse of expected Fisher information is a precise approximation of the wrong measure of precision. This implies that the FOD is optimal with respect to a less relevant measure of precision. The inverse of observed Fisher information is a higher order approximation of the conditional MSE. Designs with the objective of optimizing observed Fisher information lead to increased precision [Section \ref{sec:Eff}]. 

%There is a challenge associated incorporating observed Fisher information into the design; the observed Fisher information depends on the data and is unknown prior to the experiment. In order to incorporate observed Fisher information into the design it is required that the data can be collected sequentially, in a series of runs, 

Incorporating observed Fisher information into the design is a challenge since it depends on the data and is unknown prior to the experiment. Therefore, the method proposed in this work requires  observations to be collected sequentially, in a series of runs,  so that the observed Fisher information from the preceding runs is available to inform the design of the next run. An \textit{observed information adaptive design (OAD)} is an adaptive design with the objective of optimizing observed Fisher information  \cite{Lane:Adap:2019}. In Section \ref{sec:adapt} the theory of optimal design is used to develop an OAD for linear regression models. Linear regression is a special case of the model considered in \cite{Lane:Adap:2019} where heuristics and a simulation study were used to demonstrate the efficiency of OADs. The design proposed in Section \ref{sec:adapt} is novel; however, the main contribution of the current work are the theoretical arguments demonstrating that the proposed OAD is more efficient than the corresponding FOD with respect to conditional MSE. 

To better understand the significance of the efficiency result OADs are contrasted against traditional adaptive designs. Adaptive designs are commonly considered in experiments where the design objective depends on unknown characteristics of the population. Define a \emph{population based adaptive design (PAD)} as any procedure where some or all of the data from the preceding runs is used to update information about the population to determine the design for the next run. The vast majority of existing adaptive designs can be considered PADs; for illustrative purposes design objectives that depend on the unknown model parameters will be highlighted. 

In experiments where the design objective depends on the model parameters one common solution is to guess the parameter values and determine an efficient fixed design with respect to this guess. This can lead to very inefficient designs if the guess is far from reality. A PAD that updates the initial guess with the MLE obtained from the preceding runs is a robust alternative. Some example PADs are given below. 

Consider an experiment with several possible treatments; the design can be considered a rule that specifies how to allocate the observations to these treatments. There are many possible experimental objectives in this setting. In clinical trials a common objective is to maximize the number of subjects (observations) given the best available treatment while maintaining an adequate power to test for significant differences among the treatments. In other settings the objective is simply to assign observations to treatments in order to maximize the power of the pairwise comparisons. For both of these objectives, and others, the most efficient design depends on the underlying treatment success probabilities which can be estimated after each run with the MLE. Response adaptive randomization (RAR) designs are a popular method in this context \cite{Rose:Lach:Rand:2002}. If the success probabilities were known \emph{a priori} then it is possible to determine a fixed design that makes more efficient use of the observations and there is no need for adaptation. 

A second PAD framework is sample size reassessment (SSR) designs. In blinded SSR the data from the initial run of the experiment is used to estimate the unknown standard deviation. In unblinded SSR the initial run is used to estimate an effect size, e.g. the mean difference between two treatments divided by its standard deviation. In both blinded and unblinded SSR the estimates from the initial run are used to determine the sample size that accomplishes a pre-specified objective, e.g., the minimum sample size that guarantees a nominal power. As with RAR, if the population characteristics (standard error or effect size) were known then this minimum sample size could be determined in advance and there would be no benefit associated with an adaptive SSR design.  

The final example is particularly relevant since it is commonly proposed as a solution to the ``local" optimal design problem. Designs are \textit{locally optimal}, in a neighborhood of the true parameters, if the expected Fisher information depends on the model parameters. An adaptive optimal design (AOD) addresses the local dependence by evaluating the FOD at the MLE obtained from the preceding runs \cite{Box:Hunt:Sequ:1965,Pron:Adap:2000,Pron:Pena:2010,Pron:OneS:2010,LinFlouRose2019}. In some cases AODs have been found to perform worse than local FODs even in cases where the guess is far from the truth \cite{Dett:Born:OnTh:2013,Lane:Yao:Flou:Info:2014}. Further, the results of \cite{Hu:Rose:Opti:2003} imply, for Bernoulli outcomes, that AODs lose power at the rate $O(n^{-1})$ relative to the FOD evaluated at the true parameters. As was the case for the previous two PADs, if the parameters were known it is more efficient to use the FOD than an AOD. 

Many other PAD frameworks exist, group sequential designs, up and down designs, multi-arm bandits, etc. These designs often do not explicitly update the parameters based on the preceding runs to determine the design for the next run. However, each of these designs is characterized by a setting where the design objective depends on unknown aspects of the population and some or all of the data from the preceding runs is used to adaptively update the relevant characteristics of the population to determine the design of the next run. As was the case in the RAR, SSR and AOD examples, if the characteristics of the population relevant to the design objective were known in advance then a fixed design could be determined that is as efficient, if not more so, than the relevant PAD. In a real world experiment the parameters are never known in advance and the aforementioned adaptive design procedures are reasonable alternative to fixed designs. 

In optimal designs for linear regression models the design objective, to maximize a concave function of expected Fisher information, depends only on $f_{x}$ and $\mathcal{X}$, which, in this work, are assumed to be known. As a result the FOD is \emph{global}; in that it does not depend on any unknown characteristics of the population and it is optimal for all $\boldsymbol{\beta}\in B$. From the preceding discussion, it can be concluded that there is no need for a PAD; certainly, there is no need for an AOD. However, as previously discussed there are aspects of the observed sample that impact the precision of the MLE; specifically, the ancillary configuration matrix. A distinguishing characteristic of OADs, relative to PADs, is that they do not use the data from the preceding runs to update characteristics of the population. Specifically, an OAD does not depend on the model parameters, $\boldsymbol{\beta}$, and it is not required or useful to estimate $\boldsymbol{\beta}$ in the interim runs. Instead, OADs use the ancillary information in the \emph{observed sample} to select design points to maximize the information in the conditional distribution $\boldsymbol{\hat{\beta}}|\boldsymbol{A}$. This suggests that OADs represent a new paradigm in adaptive design where the primary objective is to accommodate new information about the observed sample and not new information about the underlying population.  

A second distinguishing feature of OADs is their relative efficiency. Population based adaptive designs (RAR, SSR, AOD, etc.) can be more robust (efficient) than fixed designs based on \emph{imperfect} knowledge of the population. However, PADs are not known to be more efficient than fixed designs based on \emph{perfect} knowledge of the population; as previously remarked, PADs are often less efficient than such designs \cite{Hu:Rose:Opti:2003,Dett:Born:OnTh:2013,Lane:Yao:Flou:Info:2014}. In other words, it is expected that there exists a fixed design that is more efficient, if not more so, than any PAD. Contrast this against the the main result in this work, stated in Section \ref{sec:Eff}, that for every fixed design, including the FOD, there exists an OAD with greater efficiency with respect to conditional MSE. Section \ref{sec:adapt} provides an algorithm to construct such an adaptive design. To the author's knowledge this is the first adaptive design shown to be more efficient than all possible fixed designs.

The efficiency proofs demonstrate a large sample benefit of the proposed OAD. A natural question is does this large sample benefit translate to finite sample benefits. The relative efficiency of the proposed design is examined in a simulation study in Section \ref{sec:sim}. The simulation study includes different error distributions, mean functions, sample sizes and optimality criteria. The proposed OAD performed nearly uniformly better than the corresponding FOD over the entire set of cases considered.

The primary motivation for the design proposed in this work is the relationship between the observed Fisher information and conditional MSE. Observed Fisher information has been more generally considered in many areas of statistical research \cite{Loui:Find:1982,Barn:Sore:ARev:1994,Murp:Van:Obse:1999,Lyst:Hugh:Exac:2002}. In adaptive design, normalizing with observed Fisher information has been found useful \cite{May:Flou:Asym:2009,LinFlouRose2019}. Therefore, there may be benefits associated with OADs in addition to optimizing conditional MSE.

\section{Information and Conditioning} \label{sec:Model}

This section assumes responses are observed from an arbitrary fixed design $\xi_{n}$. The observed and expected Fisher information measures stated in Section \ref{sec:Intro} are explicitly derived. Additionally, the extension of relevant results from \cite{Efro:Hink:Asse:1978} to linear regression, stated in equations \eqref{eq:orderI} and \eqref{eq:orderF}, are presented.

%\begin{table}
%\centering
%\scriptsize
%\begin{tabular}{ ccc }
%Notation & Brief Description & Equation Number \\
%$J_{\boldsymbol{A}}(\boldsymbol{x})$ & Observed Fisher Information & %\eqref{eq:ObsInfo2} \\
%$\mathscr{F}(\xi)$ & Expected Fisher Information  &  \\
%$M(\xi)$ & Normalized Expected Fisher Information &  \\
%$\xi_{\Psi}^{*}$ & Fixed Optimal Design (FOD) & \eqref{eq:Opt} \\
%$\check{\xi}_{\Psi}$ & Linear Regression Observed Information Adaptive %Design (ROAD) &  \\
%${\rm{MSE}}[\boldsymbol{\hat{\beta}}|\mathcal{A}]$ & Conditional MSE  \\
%${\rm{MSE}}[\boldsymbol{\hat{\beta}}]$ & Unconditional MSE  \\
%$\Psi\mbox{-Eff}_{\rm{UI}}(\xi)$ & Efficiency of the design $\xi$ relative %to $\xi_{\Psi}^{*}$ with respect to unconditional inference & %\eqref{eq:Meff}  \\
%$\Psi\mbox{-Eff}_{\rm{UMSE}}(\xi)$ & Efficiency of the design $\xi$ %relative to $\xi_{\Psi}^{*}$ with respect to unconditional MSE & %\eqref{eq:MSEEff}   \\
%$\Psi\mbox{-Eff}_{\rm{CI}}(\xi)$ & Efficiency of the design $\xi$ relative %to $\xi_{\Psi}^{*}$ with respect to conditional inference & %\eqref{eq:CIEff} \\\
%$\Psi\mbox{-Eff}_{\rm{CMSE}}(\xi)$ & Efficiency of the design $\xi$ %relative to $\xi_{\Psi}^{*}$ with respect to conditional MSE  & %\eqref{eq:CMSEEff} \\
%\end{tabular}
%\caption{Notation and a brief description of commonly used definitions and %formulas. Inline equation number is given where available.} %\label{tab:notatoin}
%\end{table}

\subsection{Information}
In Section \ref{sec:Intro} two types of information were briefly introduced; information with respect to $\eta_{i}$ and information with respect to $\boldsymbol{\beta}$. Let $l_{\eta_{i}}(y_{ij}) = \log f_{\eta}(y_{ij})$, $\dot{l}_{\eta_{i}}(y_{ij}) = (\partial/\partial\eta_{i}) l_{\eta_{i}}(y_{ij})$ and $\ddot{l}_{\eta_{i}}(y_{ij}) = (\partial^{2}/\partial\eta_{i}^{2}) l_{\eta_{i}}(y_{ij})$. From the location family assumption these terms can be written as $l_{\eta_{i}}(y_{ij}) = l_{0}(\varepsilon_{ij})$, $\dot{l}_{\eta_{i}}(y_{ij}) = \dot{l}_{0}(\varepsilon_{ij})$ and $\ddot{l}_{\eta_{i}}(y_{ij}) = \ddot{l}_{0}(\varepsilon_{ij})$.   Assuming that derivatives and integrals can be exchanged, the expected Fisher information on $\eta_{i}$ from a single observation is $\mu = -{\rm{E}}[\ddot{l}_{0}(\varepsilon_{ij})]$.
From the independent and identical nature of the errors $\mu = \mathscr{F}/n$ is constant across all $x\in\mathcal{X}$.

In linear models the information on $\boldsymbol{\beta}$ is the primary interest. From the factorization given in \eqref{eq:trans_dist}, the observed Fisher information at $\boldsymbol{\beta}'$, is defined as
\begin{align}
J_{\boldsymbol{A}}(\boldsymbol{\beta'},\boldsymbol{x}) = \sum_{i=1}^{d} \left[ \frac{\partial^{2} \log  g_{\boldsymbol{a_{i}}}[\hat{\eta}_{i} - \boldsymbol{\beta}^{T} f_{x}(x_{i})]}{\partial\boldsymbol{\beta}^{2}} \right]_{\boldsymbol{\beta} = \boldsymbol{\beta}'} \label{ObsInfoB}.
\end{align}
A traditional representation of observed Fisher information is obtained by letting $\boldsymbol{\beta}' = \hat{\boldsymbol{\beta}}$, where $\hat{\boldsymbol{\beta}}$ is the MLE of $\boldsymbol{\beta}$. Recalling that $\eta_{i}=\boldsymbol{\beta}^{T} f_{x}(x_{i})$ and it can be seen that by evaluating $\eta_{i}$ at $\hat{\eta}_{i}$ that one obtains $\boldsymbol{i_{a_{i}}}$ in the square brackets of \eqref{ObsInfoB}, for $i=1,\ldots,d$. This motivates an alternative definition of observed Fisher information 
\begin{align} \label{eq:ObsInfo2}
J_{\boldsymbol{A}}(\boldsymbol{x}) = \sum_{i=1}^{d}\boldsymbol{i_{a_{i}}} f_{x}(x_{i})f_{x}^{T}(x_{i})= F^{T} I_{\boldsymbol{A}} F.
\end{align}
The dependence of the observed Fisher information on $\boldsymbol{x}$ will be omitted when the meaning is clear. Throughout the remainder of this work any reference to observed Fisher information refers to $J_{\boldsymbol{A}}$ unless explicitly stated otherwise. As stated, $\boldsymbol{i_{a_{i}}}$ is ancillary which implies that  $J_{\boldsymbol{A}}$ is ancillary. 

Assuming derivatives and integrals can be exchanged, the expected Fisher information for model \eqref{eq:model} with design $\xi_{n}$ is $\mathcal{F}=\mathcal{F}(\xi_{n}) = E[J_{\mathcal{A}}(\boldsymbol{\beta},\boldsymbol{x})]  = \mathscr{F} F^{T}WF$. 

\subsection{Conditional Mean Square Error}
The observed Fisher information, as defined in \eqref{eq:ObsInfo2}, is not the one most commonly used in linear regression. The more common definition is $J_{\boldsymbol{A}}(\hat{\boldsymbol{\beta}},\boldsymbol{x})$. The following theorem justifies \eqref{eq:ObsInfo2} for the MLE.
\begin{theorem} \label{thm:MSE}
Under the conditions stated in Section \ref{sec:Tech} $(\boldsymbol{\hat{\beta}},\mathcal{A})$ is a sufficient statistic 
\begin{align}
{\rm{MSE}}[\boldsymbol{\hat{\beta}}|\mathcal{A}] &= J_{\mathcal{A}}^{-1}[1 + O_{p}(n^{-1})] \quad \mbox{and}\\
{\rm{MSE}}[\boldsymbol{\hat{\beta}}|\mathcal{A}] &= [\mathcal{F}(\xi_{n})]^{-1}[1 + O_{p}(n^{-1/2})].
\end{align}
\end{theorem}
Theorem \ref{thm:MSE} justifies equations \eqref{eq:orderI} and \eqref{eq:orderF}. The technical arguments for the above theorem and other main results can be found in Section \ref{sec:Tech}. 

Theorem \ref{thm:MSE} also extends the sufficiency argument for conditioning on relevant subsets. Specifically, $\boldsymbol{\hat{\beta}}$ alone is not a sufficient statistic and its distribution does not contain the full information available in the sample. Since $(\boldsymbol{\hat{\beta}},\boldsymbol{A})$ is sufficient and $\boldsymbol{A}$ is ancillary the full information is recovered in the conditional distribution $\boldsymbol{\hat{\beta}}|\boldsymbol{A}$. 

The theorem also shows that the inverse of observed Fisher information is a higher order approximation to the conditional MSE than the inverse of expected Fisher information. As stated in Section \ref{sec:Intro}, if the conditional argument, that MSE$[\hat{\boldsymbol{\beta}}|\mathcal{A}]$ is the more relevant measure of the precision is accepted, then it follows that the inverse of observed Fisher information, $J_{\boldsymbol{A}}^{-1}$, is more relevant than the inverse of expected Fisher information, $\mathcal{F}^{-1}$. 

To summarize, the primary point of this section has been to show that inference should be based on the entries of the observed Fisher information and not the entries of the expected Fisher information. This motivates the investigation in this work into designs that optimize $J_{\boldsymbol{A}}$ instead of designs that optimize $\mathcal{F}$. The discussion in this section has been under the assumption of a non-sequential experiment. In Section \ref{sec:Cond} Theorem \ref{thm:MSE} is extended to the OAD proposed in Section \ref{sec:adapt}.

\section{Optimal Designs} \label{sec:Design}

The optimal design has long been considered the benchmark of efficiency, see \cite{Kief:Opti:1959,Kief:Wolf:Opti:1959,Kief:Wolf:TheE:1960} for early foundational works. The field of optimal design continues to be an active area of statistical research; see \cite{Puke:Opti:2006} and \cite{Atki:Done:Tobi:opti:2007}, for reference texts. Here, the optimal design is referred to as the fixed optimal design (FOD); fixed, since it is obtained in advance of data collection. The term $\mathscr{F}$ is a scalar in $\mathcal{F}(\xi)$ and does not influence the optimization problem. For this reason the optimal design is commonly defined in terms of the normalized expected Fisher information $M = M(\xi_{n}) = \mathcal{F}(\xi_{n})/\mathscr{F} = F^{T}WF$. The FOD maximizes $M$, with respect to a concave criterion function, denoted $\Psi(\cdot)$. Formally, a fixed design $\xi_{\Psi}^{*}$ is $\Psi$-optimal if
\begin{align} \label{eq:Opt}
\xi_{\Psi}^{*} = \arg \max_{\xi \in \Xi} \Psi[M(\xi)],
\end{align}
where $\Xi$ represents the set of all permissible designs. The set of permissible designs has two common characterizations, exact and continuous. An optimal design is exact if the maximization in \eqref{eq:Opt} is with respect to $\Xi =\Xi_{n}$, where $\Xi_{n}$ represents the set of all possible exact designs. For continuous designs the integer restriction is relaxed to $\Xi =\Xi_{\Delta}$, where $\Xi_{\Delta}$ contains all possible designs such that $0\le w_{i}\le 1$ and $\sum_{i} w_{i} = 1$.

Let $M_{\Psi}^{*} = M(\xi_{\Psi}^{*})$ and $\Psi^{*} = \Psi(M_{\Psi}^{*})$ the normalized expected information and criterion value associated with any arbitrary, exact or continuous, FOD. A continuous FOD is  denoted $\xi_{\Psi_{\Delta}}^{*}$. In this work it is assumed that $M_{\Psi_{\Delta}}^{*} = M(\xi_{\Psi_{\Delta}}^{*})$ is positive definite and $\Psi_{\Delta}^{*} = \Psi(M_{\Psi_{\Delta}}^{*}) >0$.

In principle $\Psi$ is a non-negative positive-homogeneous concave function. In practice, $\Psi$ is selected according to the primary objective(s) of the experiment. The $D$-criterion, with $\Psi(M) = |M|^{1/p}$, minimizes the volume of the confidence ellipsoid of the MLE. The $A$-criterion defined, for non-singular $M$, as $\Psi(M) = [\mathrm{Tr}(M^{-1})]^{-1}$ minimizes the average MSE of the parameter estimates. 

An important characteristic of the linear model is that FODs do not depend on the model parameters. As a result FODs are globally optimal for all $\boldsymbol{\beta} \in B$. However, it will be shown in Section \ref{sec:Cond} that FODs are less efficient, with respect to conditional MSE, than the OAD proposed in Section \ref{sec:adapt}.

\subsection{Optimal Design Construction}

Many algorithms to construct continuous FODs are sequential. In the fixed design setting sequential means that the design points (and not the responses) from the preceding observations are available to determine the design for the next observation. The FOD can be found entirely in advance of the data collection and it is not necessary to conduct the experiment sequentially. This is different from a sequential experiment required in adaptive design, where both the design points and responses of past observations are known. %Adaptive optimal designs, described in Section \ref{sec:Intro}, have been proposed by extending sequential FOD algorithms \cite{Drag:Fedo:Adap:2005,Drag:Fedo:Wu:Adap:2007}.

In this section a general first order sequential algorithm to find continuous optimal designs is reviewed. This approach will be adapted in the next section to optimize observed Fisher information. In $D$-optimal design the first order algorithm is often referred to as the Fedorov-Wynn algorithm \cite{Wynn:TheS:1970,Fedo:Theo:1972}. The first order approach is based on the general equivalence theorem \cite{Kief:Wolf:TheE:1960}. Let $\delta_{x}$ be a design with support, $x$, and unit allocation. The \textit{sensitivity function} is defined as the derivative of $\psi = 1/\Psi$ in the direction of $\delta_{x}$; i.e., $\phi(x,\xi) = \nabla_{\delta_{x}} \psi[M(\xi)]$, where $\nabla_{\delta_{x}}$ is the derivative in the direction of $\delta_{x}$. The \textit{general equivalence theorem} states that the following are equivalent (1) $\xi_{\Psi}^{*}$ is the continuous optimal design (2) $\xi_{\Psi}^{*}=\arg\max_{\xi \in \Xi_{\Delta}}\min_{x\in\mathcal{X}}\phi(x,\xi)$ and (3) the $\min_{x\in\mathcal{X}}\phi(x,\xi_{\Psi}^{*})=0$ with equality if and only if $x$ is a support point of the FOD.

The sequential approach to FOD is initiated by selecting a design for the first $j-1$ observations, denoted $\xi(j-1)$, such that $M[\xi(j-1)]$ is non-singular. The general equivalence theorem implies that if $\xi(j-1)$ is any arbitrary non-optimal design then the minimum of $\phi[x,\xi(j-1)]<0$.  Let
$\xi(j) = \alpha_{j-1}\delta_{x} + (1-\alpha_{j-1})\xi(j-1)$, where $\alpha_{j-1}\in[0,1]$ is the step size of the search algorithm; then the expected Fisher information can be written as
\begin{align} \label{eq:SeqM1}
M[\xi(j)] = (1-\alpha_{j-1}) M[\xi(j-1)] + \alpha_{j-1} M(\delta_{x}).
\end{align}
The general equivalence theorem guarantees the existence of a
\begin{align} \label{eq:FODxj}
\overline{x}(j) = \min_{x\in\mathcal{X}} \phi[x,\xi(j-1)]
\end{align}
such that $\Psi[\xi(j)]>\Psi[\xi(j-1)]$. To obtain the optimal design this procedure is iterated for $j=1,\ldots,D$. See \cite{Wu:Wynn:1978} for a discussion of when $\lim_{D\rightarrow\infty}\Psi\{M[\xi(D)]\} = \Psi_{\Delta}^{*}$. 
 
It is also possible to search for the $\alpha_{j-1}$ that maximizes $\Psi\{M[\xi(j)]\}$ \cite{Fedo:Theo:1972}, but this is not desired in the context of an adaptive experiment.  Selecting $\alpha_{j-1}= 1/j$ mimics a fully sequential experiment where the $j$th observation is placed at $\overline{x}(j)$. 

\section{Adaptive Design} \label{sec:adapt}
Observed information adaptive design (OAD) has been considered in a more general setting where the observed and expected Fisher information depend on the model parameters \cite{Lane:Adap:2019}. Two OADs, the local observed information adaptive design (LOAD) and the maximum likelihood estimated observed information adaptive design (MOAD) were proposed to address the local dependence. As shown in Section \ref{sec:Model},  for the linear model, observed and expected Fisher information do not have a local dependence on the model parameters. The design proposed in this section is not equivalent to either the LOAD or MOAD procedure.

\subsection{Sequential Experiment}
In the sequential setting the sample is an ordered set of experimental runs. In the fully sequentially setting each run is comprised of a single observation. Specifically, the data for observations $1,\ldots,j-1$ are known prior to the design assignment of the $j$th observation, for $j=2,\dots,n$.  The ancillary configuration statistic for the $i$th design point from the first $j$ runs is denoted $\boldsymbol{a}_{i}(j)$ and the corresponding ancillary configuration matrix is denoted $\boldsymbol{A}(j) = [\boldsymbol{a}_{1}(j),\ldots,\boldsymbol{a}_{d}(j)]$.

Before proceeding, some notation relevant to the sequential observed Fisher information is needed. Let $q_{\boldsymbol{a}_{i}}(j) = \boldsymbol{i}_{\boldsymbol{a}_{i}(j)}/\mu$, $Q_{\boldsymbol{A}(j)} = \sum_{i} q_{\boldsymbol{a}_{i}(j)}$ and 
\begin{align}
\tau_{\boldsymbol{A}(j)} = \left\{\begin{array}{cccc} x_{1} & x_{2} & \ldots & x_{d} \\ \omega_{\boldsymbol{a}_{1}(j)} & \omega_{\boldsymbol{a}_{2}(j)} & \ldots & \omega_{\boldsymbol{a}_{d}(j)} \end{array}\right\},
\end{align}
where $\omega_{\boldsymbol{a}_{i}(j)} = q_{\boldsymbol{a}_{i}(j)}/Q_{\boldsymbol{A}(j)}$. As stated in \cite{Barn:Sore:ARev:1994} $\boldsymbol{i}_{\boldsymbol{a}_{i}(j)}\ge 0$ which implies $q_{\boldsymbol{a}_{i}(j)}\ge 0$; however, there could exist positive probability that it equals zero. This is unlikely to occur in practice and setting $q_{\boldsymbol{a}_{i}(j)}$ to be equal to a small positive constant when this occurs ensures that $\tau_{\boldsymbol{A}(j)}$ is a continuous design, i.e. $\tau_{\boldsymbol{A}(j)}\in \Xi_{\Delta}$. As shown in \cite{Lane:Adap:2019}, the observed Fisher information from the first $j$ observations can be written as
\begin{align}
J_{\boldsymbol{A}(j)}(\boldsymbol{x}) = Q_{\boldsymbol{A}(j)} M[\tau_{\boldsymbol{A}(j)}].
\end{align}
The above indicates that, after run $j$, $J_{\boldsymbol{A}(j)}$ is proportional, up to a known constant, to the expected Fisher information evaluated at the design $\tau_{\boldsymbol{A}(j)}$. Each of the quantities introduced in this section are functions of only the ancillary configuration statistics and are themselves ancillary.

\subsection{Observed Information Adaptive Design for Linear Regression}
To construct the adaptive design proposed in this section it is required that the a FOD, with respect to $\Psi$, denoted $\xi_{\Psi}^{*}$, is known or can be computed. It is not required that the FOD is found using the first order sequential approach. It can be found using any fixed or continuous FOD approach that results in a $\xi_{\Psi}^{*}$ such that $\Psi[M(\xi_{\Psi}^{*})] = \Psi_{\Delta}^{*}[1+O(n^{-1})]$.  Recall, the support points of the FOD and their corresponding allocations are denoted $\boldsymbol{x}^{*} = (x_{1}^{*},\ldots,x_{d}^{*})^{T}$ and $\boldsymbol{w}^{*} = (w_{1}^{*},\ldots,w_{d}^{*})^{T}$, respectively. %The proposed OAD, defined in this section, sequentially allocates observations to the support points of the FOD; non-optimal design points will not be considered.

The objective of an OAD is to optimize the observed Fisher information, $J_{\boldsymbol{A}(n)}$, with respect to a concave criterion. A challenge with optimizing $J_{\boldsymbol{A}(n)}$ is that it is unknown prior to data collection. However, after $j-1$ observations $J_{\boldsymbol{A}(j-1)}$ is known. An intuitive objective is to select the support point for observation $j$ that optimizes $J_{\boldsymbol{A}(j)}$. This too is not generally possible since $J_{\boldsymbol{A}(j)}$ is unknown based on only the data from the first $j-1$ observations. A reasonable approximation for this quantity is the sum of the observed Fisher information from the preceding observations plus the expected Fisher information from one additional observation with support $x$, i.e.,
\begin{align} 
    J_{\boldsymbol{A}(j)} \approx J_{\boldsymbol{A}(j-1)} + M(\delta_{x}) 
    \propto (1-\beta_{j})M[\tau_{\boldsymbol{A}(j-1)}] - \beta_{j}M(\delta_{x}), \label{eq:Japprox}
\end{align}   
where $\beta_{j} = [1 + Q_{\boldsymbol{A}(j-1)}]^{-1}$. The omitted proportional constant in the above, $\beta_{j}^{-1}$, is known after run $j-1$. The approximation in the first line of \eqref{eq:Japprox} is valid for large $j$. Substituting $\beta_{j}$ for $\alpha_{j}$ and $\tau_{\boldsymbol{A}(j-1)}$ for $\xi(j-1)$ in \eqref{eq:SeqM1} and it can be seen that \eqref{eq:Japprox} has the same basic form as \eqref{eq:SeqM1}. Further, as previously stated, $\tau_{\boldsymbol{A}(j-1)}\in \Xi_{\Delta}$ is a design. Therefore, the first order FOD algorithm can be directly implemented to optimize the right hand side of \eqref{eq:Japprox} by replacing $\xi(j-1)$ with $\tau_{\boldsymbol{A}(j-1)}$ in equation \eqref{eq:FODxj}.

Even though the search for the design point that minimizes the sensitivity function could be carried out using a first order FOD algorithm; here a slightly different approach is employed. The support points of the FOD are known; therefore, the search for the design point for the $j$th observation need only be over $\boldsymbol{x}^{*}$. In fact, if $\omega_{\boldsymbol{a}_{i}(j-1)}>w_{i}^{*}$ then it indicates that the current observed allocation to $x_{i}^{*}$ exceeds the optimal allocation and the search can be improved by excluding such points. This motivates defining the design for the $j$th observation as
\begin{align} \label{eq:xjcheck}
\check{x}(j) = \min_{x\in\boldsymbol{x}_{+}^{*}(j)} \phi[x,\tau_{\boldsymbol{A}(j-1)}].
\end{align}
where $\boldsymbol{x}_{+}^{*}(j) = \{x_{i}^{*}:\omega_{\boldsymbol{a}_{i}(j-1)}<w_{i}^{*}, i =1,\ldots,d\}$. The following algorithm is the OAD proposed for linear regression models, referred to as the ROAD for the remainder of this work.

\begin{algorithm} Linear \underline{R}egression \underline{O}bserved Information \underline{A}daptive \underline{D}esign (ROAD) 
\begin{enumerate}[nolistsep]
\item{Find a FOD $\xi_{\Psi}^{*}$. For observations $j=1,\ldots,kd$ initiate the design by placing a finite number, $k$, observations on each of the support points of $\xi_{\Psi}^{*}$.}
\item{The design point for observation $j=kd+1,\ldots,n$ is $\check{x}(j)$ as defined in \eqref{eq:xjcheck}. Let $\check{\xi}_{\Psi} = \sum_{j=1}^{n}\delta_{\check{x}(j)}$ represent the observed design resulting design from a sample of size $n$.}
\end{enumerate}
\end{algorithm}

For certain distributions if $k$ is too small then the observed Fisher information may be volatile. The value $k$ should be selected to reduce this volatility. 

A natural question is whether or not only the points $\boldsymbol{x}^{*}$ should be considered in the ROAD. This restriction is required for the technical arguments in this work. Replacing $\boldsymbol{x}_{+}^{*}(j)$ with $\mathcal{X}$ in \eqref{eq:xjcheck} could potentially result in designs that are more efficient; however, it is not trivial to extend the results to this design.

\section{Conditional Inference} \label{sec:Cond}
This section address the four challenges associated with conditioning following an adaptive design identified in Section \ref{sec:Intro}. Before proceeding, some additional notation is required. For the remainder of this work statistics associated with a ROAD will be represented with a check, e.g. $\check{\boldsymbol{y}}$ and $\check{\boldsymbol{A}}$ represent the responses and the ancillary configuration matrix following a ROAD, respectively. Statistics accented with a star correspond to a FOD, eg. $\boldsymbol{y^{*}}$ and $\boldsymbol{A^{*}}$. Statistics without an accent are arbitrary and can represent the ROAD, the FOD or any other fixed design. Additionally, let $\check{\boldsymbol{X}} = (\check{x}_{1},\ldots,\check{x}_{n})$ and $\boldsymbol{X} = (x_{1},\ldots,x_{n})$ represent the sequence of observed designs points for a ROAD and an arbitrary fixed design, respectively. Further, following the ROAD it is required to condition on the filtration $\mathscr{\check{A}} = \sigma[\mathcal{\check{A}}(1),\ldots,\mathcal{\check{A}}(n)]$, which is the sub-$\sigma$ algebra generated by the random ancillary configuration matrices from all runs. Note, random variables and statistics corresponding to the  entire sample size will often have their dependence on $n$ omitted, e.g.  $\mathcal{\check{A}} = \mathcal{\check{A}}(n)$.

The first result of this section considers the conditional distribution of the responses following the ROAD.
\begin{theorem} \label{thm:ancillary}
Under the conditions stated in Section \ref{sec:Tech} $\mathscr{\check{A}}$ and $\check{\boldsymbol{X}}$ are ancillary random variables, $\check{\boldsymbol{X}}$ is $\mathscr{\check{A}}$ measurable and
\begin{align} \label{eq:eqiv_dist}
\boldsymbol{\check{Y}}|\mathscr{\check{A}} \stackrel{d}{=} \boldsymbol{Y}|\{\boldsymbol{X} = \check{\boldsymbol{X}},\mathcal{A} = \mathcal{\check{A}}\},
\end{align}
where $\stackrel{d}{=}$ represents equality in distribution.
\end{theorem}
This theorem address the first and second challenges of conditioning in adaptive designs. The design matrix is ancillary following a ROAD and there is no information lost by conditioning. This theorem also states that conditional on $\mathscr{\check{A}}$ the responses of the ROAD have the same distribution as the responses from a fixed design with the same sequence of design points, $\boldsymbol{X} = \check{\boldsymbol{X}}$, and the same ancillary configuration matrix, $\mathcal{A} = \mathcal{\check{A}}$. The implication is that with respect to conditional inference the adaptive nature of the design can be ignored. The MLE is a function of the responses and thus the above result applies, i.e. $\hat{\boldsymbol{\beta}}|\mathscr{\check{A}} \stackrel{d}{=} \hat{\boldsymbol{\beta}}|\{\boldsymbol{X} = \check{\boldsymbol{X}},\mathcal{A} = \mathcal{\check{A}}\}$. Taking for granted that conditional inference is preferred to unconditional inference implies that inference can be conducted after a ROAD as if no adaptation had taken place. Theorem \ref{thm:ancillary} does not extend to the unconditional distribution. This is a consequence of the induced dependence between $\mathcal{\check{A}}(i)$ and $\mathcal{\check{A}}(j)$, for $i,j=1,\ldots,n$.

The following corollary gives the conditional asymptotic properties of the ROAD.
\begin{corollary}\label{cor:efro:hink}
Under the conditions stated in Section \ref{sec:Tech} $(\hat{\boldsymbol{\beta}},\mathscr{\check{A}})$ is a sufficient statistic,
\begin{align}
{\rm{MSE}}[\boldsymbol{\hat{\beta}}|\mathscr{\check{A}}] &= J_{\mathcal{\check{A}}}^{-1}[1 + O_{p}(n^{-1})] \quad \mbox{and} \\
{\rm{MSE}}[\boldsymbol{\hat{\beta}}|\mathscr{\check{A}}] &= [\mathcal{F}(\check{\xi}_{\Psi})]^{-1}[1 + O_{p}(n^{-1/2})].
\end{align}
\end{corollary}
The proof of this and other corollaries are supplied in the supplemental materials. This corollary extends the higher order property of the observed Fisher information given in Theorem \ref{thm:MSE} to the ROAD. This solves the third and fourth challenge associated with conditional inference in adaptive designs. Since  $(\boldsymbol{\hat{\beta}},\mathscr{\check{A}})$ is a sufficient statistic and $\mathscr{\check{A}}$ is ancillary the arguments for conditioning on $\mathscr{\check{A}}$ are exactly analogous to those presented in Section \ref{sec:Model} regarding conditioning on $\mathcal{A}$. Further, the argument for using observed Fisher information to approximate the conditional MSE is just as strong as if no adaptation had taken place. Therefore, large sample inference after a ROAD can be conducted using the observed Fisher information with the same order of accuracy, $O_{p}(n^{-1})$, present in fixed experiments.

\section{Efficiency of the Linear Regression Observed Information Adaptive Design} \label{sec:Eff}

An alternative way to characterize the FOD is to say that it is efficient. The $\Psi$-efficiency of an arbitrary design, $\xi$, is denoted
\begin{align} \label{eq:Meff}
   \Psi\mbox{-Eff}_{\rm{UI}}(\xi) = \frac{\Psi[M(\xi)]}{\Psi_{\Delta}^{*}},
\end{align}
where $\Psi\mbox{-Eff}_{\rm{UI}}(\xi)\le1$ for all $\xi\in\Xi_{\Delta}$. A design, $\xi$, is a continuous $\Psi$-optimal design if an only if it is efficient, i.e. $\Psi\mbox{-Eff}_{\rm{UI}}(\xi)=1$. This is the definition of efficiency most commonly used in the optimal design literature, see \cite{Atki:Done:Tobi:opti:2007}. In this work, this definition of efficiency is best understood as a measure of the efficiency with respect to unconditional inference. The expected Fisher information is a second order approximation to the unconditional MSE. This property justifies the use of unconditional confidence regions and test statistics based on the entries of $n\mu M = \mathcal{F}$. The subscript UI references that $\Psi\mbox{-Eff}_{\rm{UI}}(\xi)$ describes the efficiency of a design $\xi$, relative to $\xi_{\Psi_{\Delta}}^{*}$, with respect to unconditional inference. As argued in Section \ref{sec:Model} inference based entries of $J_{\boldsymbol{\check{A}}}$ is superior to inference based on $\mathcal{F}$. In this section an analogous measure of efficiency with respect to conditional inference is developed.

Alternatively, $\Psi\mbox{-Eff}_{\rm{UI}}$ can be viewed as an approximation to the relative efficiency with respect to unconditional MSE defined as
\begin{align} \label{eq:MSEeff}
    \Psi\mbox{-Eff}_{\rm{UMSE}}(\xi) = \frac{\Psi\{{\rm{MSE}}[\hat{\boldsymbol{\beta}}(\xi)]^{-1}\}}{\Psi\{{\rm{MSE}}[\hat{\boldsymbol{\beta}}(\xi_{\Psi}^{*})]^{-1}\}},
\end{align} 
where $\hat{\boldsymbol{\beta}}(\xi)$ denotes the MLE from an experiment with design $\xi$. Note $\Psi\mbox{-Eff}_{\rm{UMSE}}$ is not bound above by 1. The measure $\Psi\mbox{-Eff}_{\rm{UMSE}}$ describes the efficiency of the design $\xi$, relative to $\xi_{\Psi}^{*}$, with respect to unconditional MSE. From the conditional perspective  $\Psi\mbox{-Eff}_{\rm{UMSE}}$ is a sub-optimal measure of efficiency. In this section a measure of efficiency with respect to conditional MSE is presented. The main result of this work, that the ROAD has greater conditional efficiency than the FOD, is stated at the end of this section.

\subsection{Conditional Inference Efficiency}

In Section \ref{sec:Model} it was argued that inference for a fixed design should be based on the entries of $J_{\boldsymbol{A}}$. Theorem \ref{thm:ancillary} established that the ROAD can be analyzed exactly as a fixed design without a loss of information. Therefore, the ROAD should be analyzed using the entries of $J_{\boldsymbol{\check{A}}}$. For example a confidence ellipsoid, for $\boldsymbol{\beta}$, can be defined as the interior of
\begin{align} \label{eq:Volume}
(\hat{\boldsymbol{\beta}} -  \boldsymbol{\beta})^{T}J_{\boldsymbol{A}}(\hat{\boldsymbol{\beta}} -  \boldsymbol{\beta}) = \chi_{p}^{2(1-\alpha)},
\end{align}
where $\chi_{p}^{2(1-\alpha)}$ is the $1-\alpha$ quantile of an $\chi^{2}_{p}$-distribution. This confidence ellipsoid is proportional to the inverse of the square root of the determinant of $J_{\boldsymbol{A}}$. A ROAD for the $D$-optimal criterion should minimize this volume. 

Accepting that $J_{\boldsymbol{A}}$ should be used to define confidence regions or test statistics implies that the efficiency of a design with respect to this measure is of interest. Note $J_{\boldsymbol{A}}$ relates to post-data inference. Prior to an experiment it is desired to ensure a design scheme improves the expected conditional inference. The measure
\begin{align} \label{eq:CIEff}
    \Psi\mbox{-Eff}_{\rm{CI}} = \frac{E[\Psi(J_{\mathcal{\check{A}}})]}{E[\Psi(J_{\mathcal{A^{*}}})]},
\end{align}
represents the efficiency of the ROAD relative to $\xi_{\Psi}^{*}$ with respect to conditional inference. If $\Psi\mbox{-Eff}_{\rm{CI}}>1$ then the ROAD is more efficient than a FOD with respect to this measure.

Before the main efficiency results are presented certain quantities that impact the relative expected efficiency need to be introduced. The first is the statistical curvature which describes the asymptotic difference between the observed and expected Fisher information \cite{Efro:Hink:Asse:1978}. \textit{Statistical curvature} is defined as
\begin{align} \label{eq:gamma}
\gamma = \left(\nu_{02}\nu_{20} - \nu_{11}/\nu_{20}^{3}\right)^{1/2},
\end{align}
where $\nu_{kl} = {\rm{E}}[\dot{l}_{0}^{k}(\varepsilon_{ij})(\ddot{l}_{0}(\varepsilon_{ij}) + {\rm{E}}[\dot{l}_{0}(\varepsilon_{ij})]^{2})^{l} ]$ \cite{Efro:Defi:1975}. In the location family $\mu = \nu_{20}$ and $\gamma$ do not depend on $\eta$, and $\gamma$ is invariant under monotonic transformations. In other words the curvature with respect to $y$ is equivalent to the curvature with respect to $\varepsilon$; this ensures that statistical curvature is constant across the design region. Intuitively, it is expected that the difference in efficiency between the ROAD and the FOD is a function of the statistical curvature. This intuition is be proven correct in Theorem \ref{thm:opt}. Note, $\gamma=0$ if the errors are normally distributed.

A second form of curvature affecting the efficiency of the ROAD and FOD is the \textit{design curvature}. Design curvature is defined as the negative Hessian matrix of $\Psi$, with respect to $\boldsymbol{w}_{(d)}^{*}=(w_{1}^{*},\ldots,w_{d-1}^{*})^{T}$ with $\boldsymbol{x}^{*}$ held fixed, evaluated at the FOD; i.e.
$H_{\Psi}^{*} = -\left.\nabla^{2}{\Psi}(M)\right|_{M = M_{\Psi}^{*}}$. The Hessian matrix of a concave optimality criterion evaluated at its maximum is negative definite and describes how small changes in the allocation weights away from the optimal allocations impact efficiency. Here $H_{\Psi}^{*}$ is defined to be positive definite for clarity. Intuitively it would be expected that a large design curvature will result in the ROAD having greater relative efficiency. The Hessian matrix $H_{\Psi}^{*}=0$ if $d=1$.

The following theorem shows how the efficiency of the ROAD and FOD change in the presence of statistical and design curvature.
\begin{theorem} \label{thm:opt}
Let $\xi_{\Psi}^{*}$ be the optimal design used in Step 1 of Algorithm 1 and $J_{\boldsymbol{A^{*}}}$ be the observed Fisher information from a fixed design with design $\xi_{\Psi}^{*}$. Under the conditions stated in Section \ref{sec:Tech}
\begin{align}
{\rm{E}}[\Psi(J_{\mathcal{A^{*}}})] &= h\Psi^{*}(n - \gamma^{2}R_{\Psi}^{*}) + o(1) \quad \mbox{and} \\
{\rm{E}}[\Psi(J_{\mathcal{\check{A}}})] &= h\Psi^{*}(n) + o(1),
\end{align}
where 
\begin{align}
R_{\Psi}^{*} =\frac{1}{2\Psi^{*}}{\rm{tr}}(H_{\Psi}^{*}V_{\Psi}^{*}), \quad h = 1 +  \frac{d}{2n\mu^{2}}\left(\frac{\mu_{3}^{2}}{\mu} + \mu_{4}\right),
\end{align} 
$V_{\Psi}^{*} =\dot{\boldsymbol{g}}(\boldsymbol{w}^{*})W^{*}\dot{\boldsymbol{g}}^{T}(\boldsymbol{w}^{*})$, $\mu_{k} = E[(\partial^{k} / \partial \eta^{k})l_{\eta}(y)]$, $\dot{\boldsymbol{g}}(\boldsymbol{w}^{*}) = (I_{d-1} - J_{\Psi}^{*},w_{d}^{*}\boldsymbol{1}_{d-1}^{T})$, $J_{\Psi}^{*} = J_{d-1}diag(\boldsymbol{w}_{(d)}^{*})$, $I_{d-1}$ is the identity matrix and $J_{d-1}$ is a $(d-1)\times(d-1)$ matrix where each element is equal to 1.
\end{theorem}
The interpretation of the this theorem is as follows; suppose an experiment is conducted with fixed design $\xi_{\Psi}^{*}$ and sample size $n$. It is expected, for large $n$, that volume of the confidence ellipsoid / power of this experiment is equal to a ROAD with a sample size of $n -  \gamma^{2}R_{\Psi}^{*}$. In other words a ROAD requires $\gamma^{2}R_{\Psi}^{*}$ fewer observations to generate confidence regions with equivalent area/volume. 

An immediate corollary of this theorem demonstrates the relative efficiency of the ROAD with respect to conditional inference.
\begin{corollary} \label{cor:opt}
Under the conditions and notation of Theorem \ref{thm:opt}
$\Psi{\mbox{\emph{-Eff}}}_{\rm{CI}} = S_{\Psi}^{*} + o(1)$, where $S_{\Psi}^{*} = (1 - \gamma^{2}R_{\Psi}^{*}/n)^{-1} > 1$ if $\gamma>0$ and $d>1$.
\end{corollary}

This corollary establishes that if a model has non-zero statistical curvature, $\gamma>0$, and the optimal design $\xi_{\Psi}^{*}$ has positive weight on more than one support point, $d>1$, then the ROAD has greater expected efficiency, with respect to conditional inference, than the corresponding FOD. This will lead to narrower confidence regions or greater power depending on the nature of the optimality criterion. As stated, $\gamma=0$ for normally distributed errors and thus $\Psi{\mbox{-Eff}}_{\rm{CI}}\rightarrow 1$ in this case.

The theorem and its corollary also provide insight into when the improvement in the relative efficiency of the ROAD can be expected to be significant. Clearly, as the statistical curvature, $\gamma$, increases the relative benefit of the ROAD will increase. The features of the design and optimality criteria also change the relative benefit, through $R_{\Psi}^{*}$; however, this relationship is more complex and is examined in the context of several examples Section  \ref{sec:ex}.

\subsection{Conditional Mean Square Error Efficiency}

Theorem \ref{thm:opt} and Corollary \ref{cor:opt} ensure that the ROAD optimizes the conditional confidence regions and power. However, this does not necessarily equate to greater precision of the MLE. The precision of the MLE is best described by the conditional MSE. In this section the preceding results are extended to the conditional MSE. Note, conditional MSE and observed Fisher information are inversely related. Therefore, the design that is optimal with respect to the conditional MSE is defined as the design that maximizes $\Psi\{{\rm{MSE}}[\hat{\boldsymbol{\beta}}|\mathcal{A}]^{-1}\}$. The conditional MSE is a random variable so when planning an experiment the expected efficiency of the ROAD relative to the FOD, defined as 
\begin{align} \label{eq:CMSEEff}
    \Psi\mbox{-Eff}_{\rm{CMSE}} = \frac{{\rm{E}}[\Psi({\rm{MSE}}[\hat{\boldsymbol{\beta}}|\mathscr{\check{A}}]^{-1})]}{{\rm{E}}[\Psi({\rm{MSE}}[\hat{\boldsymbol{\beta}}|\mathcal{A^{*}}]^{-1})]},
\end{align}
is of interest. The measure $\Psi\mbox{-Eff}_{\rm{CMSE}}$ represents the efficiency of the ROAD relative to $\xi_{\Psi}^{*}$, with respect to conditional MSE. If it is accepted that conditional MSE is a more appropriate measure of precision than the unconditional MSE then it is implied that $\Psi\mbox{-Eff}_{\rm{CMSE}}$ is a more appropriate measure of efficiency than $\Psi\mbox{-Eff}_{\rm{UMSE}}$. 

The below theorem states that the ROAD has greater expected conditional MSE efficiency than the FOD.
\begin{theorem} \label{thm:MSEopt}
Let $\xi_{\Psi}^{*}$ be the optimal design used in Step 1 of Algorithm 1 and $\mathcal{A^{*}}$ be the ancillary configuration matrix from a fixed design with design $\xi_{\Psi}^{*}$. Under the conditions stated in Section \ref{sec:Tech}
\begin{align}
{\rm{E}}[\Psi({\rm{MSE}}[\hat{\boldsymbol{\beta}}|\mathscr{\check{A}}] - \Psi({\rm{MSE}}[\hat{\boldsymbol{\beta}}|\mathcal{A^{*}}]^{-1})] = \gamma^{2}tr(H_{\Psi}^{*}V_{\Psi}^{*})/2 + o(1), \label{eq:MSEEff}
\end{align}
where $tr(H_{\Psi}^{*}V_{\Psi}^{*}) >1$ if $\gamma>0$ and $d>1$.
\end{theorem}
The difference in efficiency given in \eqref{eq:MSEEff} is, perhaps, the clearest result in favor of the ROAD. It implies that $\Psi\mbox{-Eff}_{\rm{CMSE}}$ is greater than 1 as $n\rightarrow\infty$ which confirms that the ROAD is more efficient with respect to expected conditional MSE than the FOD, for large $n$. Another way of interpreting the theorem is to say that for every FOD [continuous or exact up to order $O(n^{-1})$] using the ROAD \emph{always} increases the asymptotic efficiency with respect to conditional MSE. 

\begin{remark}
Theorems \ref{thm:opt} and \ref{thm:MSEopt} established that any design such that $\Psi[M(\xi_{\Psi}^{*})] = \Psi_{\Delta}^{*}[1+O(n^{-1}]$ is less efficient than the corresponding ROAD. Consider any design $\xi$ such that $\Psi[M(\xi)] = \Psi_{\Delta}^{*}(1+cn^{-\delta})$, where $c>0$ and $\delta\in[0,1)$ are fixed constants. Let $\mathcal{A}$ correspond to the ancillary configuration matrix from the design $\xi$. In the ntal materials it is shown that ${\rm{E}}[\Psi({\rm{MSE}}[\hat{\boldsymbol{\beta}}|\mathscr{\check{A}}] - \Psi({\rm{MSE}}[\hat{\boldsymbol{\beta}}|\mathscr{A}]^{-1})]\rightarrow\infty$ as $n\rightarrow\infty$,  This remark in combination with Theorem \ref{thm:MSEopt} proves that for every fixed design there exists a ROAD that is more efficient with respect to conditional MSE. In other words in order to maximize the precision of the MLE \emph{requires} adaptation. In the supplement it is also shown that ${\rm{E}}[\Psi(J_{\mathcal{\check{A}}})] - {\rm{E}}[\Psi(J_{\mathcal{A}})] \rightarrow \infty$, as $n\rightarrow\infty$ so a similar statement holds for the efficiency of the ROAD with respect to conditional inference. 
\end{remark}

The results of this section have shown that by considering a conditional approach in the design of the experiment it is possible to see expected benefits.  The expected benefit is important when planning an experiment. For example, it is possible to approximate how many fewer observations are required using a ROAD in place of a FOD in order to have the same expected volume of a confidence ellipsoid. However, one should not lose sight of the conditional benefits. Once an experiment is completed the average benefit is irrelevant. It is of greater interest to know what is the observed volume of the confidence ellipsoid found using \eqref{eq:Volume} or the observed conditional MSE. The conditional benefits of the ROAD are, likely, greater than the unconditional benefits; however, since conditional benefits are only relevant to the observed experiment they are difficult to quantify and examine.

Remark, $\Psi\mbox{-Eff}_{\rm{CMSE}}$ is a useful tool to demonstrate a theoretical benefit in term of conditional MSE. Practically, this measure is unlikely to be able to be computed since in most cases there is no known closed form expression available for the conditional MSE.

\section{Illustrative Examples} \label{sec:ex}

Theorems \ref{thm:opt} and \ref{thm:MSEopt} indicate that a ROAD with sample size $n$ has the same efficiency as a FOD with a sample of size $n+\gamma^{2}R_{\Psi}^{*}$. In other words the number of samples ``saved'' by using a ROAD is $\gamma^{2}R_{\Psi}^{*}$. For certain error distributions  $\gamma^{2}$  can be made to be arbitrarily large \cite{Efro:Defi:1975}. Statistical curvature is given for some examples error distributions in Section \ref{sec:sim}. In this section $R_{\Psi}^{*}$ is examined for three different models and three different optimality criteria. 

\emph{Treatment Model}: Consider an experiment with $s$ treatments, indexed by $1,\ldots,s$, where each subject can receive only 1 treatment. This can be modeled with $f_{x}(x) = (x_{1},\ldots,x_{s})^{T}$, where $x_{i} = 1$ for the treatment $i$ and 0 otherwise, $i=1,\dots,s$. The design region in the treatment model is the collection of vectors indicating each individual treatment. 

\emph{Interaction Model}: The second example extends the treatment model to include all two-way interactions. For this model $f_{x}(x) = (x_{1},\ldots,x_{s},x_{1}x_{2}\ldots,x_{s-1}x_{s})^{T}$, with $x_{i}$ defined as before. The design region consists of the collection of vectors indicating at most two treatments. 

\emph{Quadratic Model}: The final model considered is a second order polynomial with $s$ quantitative covariates and $f_{x}(x) = (x_{1},\ldots,x_{s},x_{1}^{2},\ldots,x_{s}^{2},x_{1}x_{2}\ldots,x_{s-1}x_{s})^{T}$. The design region considered for this example is $\mathcal{\boldsymbol{X}} = [0,1]^{s}$. 

For the treatment and interaction model it can be verified, using the general equivalence theorem, that the continuous $D$-optimal design places equal weight on each point in $\mathcal{X}$. The quadratic example is more complicated. The support points of the $D$-optimal design are a subset of the $3^{r}$ support points of a full factorial design with levels 0, 1/2 and 1 \cite{Farr:Kief:Walb:Opti:1968}. For the treatment model the $A$- and $D$-optimal designs are equivalent. The $A$-optimal designs for the interaction and quadratic models were found using the R package \textbf{OptimalDesign} \cite{Filo:Harm:Opti:2016}.

Table \ref{tab:LargeSample} presents $R_{\Psi}^{*}$ for the treatment, interaction and quadratic models for $s=1,\ldots,9$. The results for the $D$- and $A$-optimal design appear in the columns $D$ and $A$, respectively. The number of parameters, $p$, for each model is also given. Recall, from the discussion following Theorem \ref{thm:opt} that $\gamma^{2}R_{\Psi}^{*}$ represents the number of additional samples required for the FOD to have the same efficiency as the ROAD with $n$ samples. As an example of how to read this table consider the quadratic model with 4 treatments. For this example fixed $D$-optimal design requires $13.0\gamma^{2}$ additional observations in order to have the same efficiency as the ROAD. The sample size saving can be considerable; the most extreme example is for the quadratic model with $s=9$ where $R_{A}^{*}=1312.2$. Of course it would require a very large sample size for this benefit to be realized.  

A remark on $R_{D}^{*}$; this term is presented for the $D$-optimality criteria defined as $|M|^{1/p}$. However, this is not the only characterization of the criteria; $\log(M)$ is also commonly used. The value $R_{D}^{*}$ will vary by the definition of the criteria. When comparing the ROAD to the FOD it is important that you compare with respect to the scale of interest. %For example in Section \ref{sec:Eff} it was stated that the volume of the confidence ellipsoid is proportional to $|J_{\boldsymbol{A}}|^{1/2}$. Therefore, if it is of interest to determine the sample size of a ROAD that will lead to the same volume of the confidence ellipsoid from a FOD with size $n$. If the notation for the $D$-criterion, $D(M) = |M|^{1/p}$, is used then $[D(M)]^{p/2} = |M|^{1/2}$ represents the scale of the volume of the confidence ellipsoid. Let $D^{p/2}$ represent the preceding characterization of the criterion; then  $R_{D^{p/2}}^{*} = \frac{p}{2}R_{D}^{*}$. This illustrates that on this scale you must multiple the values in the Table \ref{tab:LargeSample} by $p/2$ to represent the sample size savings on the scale of the volume of the confidence ellipsoid. For example, for the quadratic model with $s=9$ the sample size savings was $53.8\gamma^{2}$ for $D = |M|^{1/p}$; however, for $D^{p/2} = [D(M)]^{p/2} = |M|^{1/2}$ the sample size saving is $55*53.8\gamma^{2}/2 = 12,523.5\gamma^{2}$. 

\begin{table}
\centering
\setlength\tabcolsep{5.0pt}
\renewcommand\arraystretch{1.1}
\scriptsize
\begin{tabular}{cccccccccc}
\hline
& \multicolumn{3}{c}{Treatment} & \multicolumn{3}{c}{Interaction} & \multicolumn{3}{c}{Quadratic}\\
\hline
$s$ & $p$ & $D$ & $A$ & $p$ & $D$ & $A$ &  $p$ & $D$ & $A$  \\
\hline
1 & 1 & 0.0 & 0.0 & 2 & 1.5 & 1.5 & 3 & 1.0 & 2.0 \\
2 & 2 & 0.5 & 0.5 & 4 & 1.5 & 1.5 & 6 & 2.0 & 21.7 \\
3 & 3 & 1.0 & 1.0 & 7 & 5.1 & 5.8 & 10 & 4.1 & 25.4 \\
4 & 4 & 1.5 & 1.5 & 11 & 5.0 & 12.3 & 15 & 13.0 & 110.8 \\
5 & 5 & 2.0 & 2.0 & 16 & 7.5 & 24.7 & 21 & 56.3 & 309.5 \\
6 & 6 & 2.5 & 2.5 & 22 & 10.5 & 41.1 & 28 & 72.8 & 512.7 \\
7 & 7 & 3.0 & 3.0 & 29 & 14.0 & 63.5 & 36 & 165.0 & 1077.4 \\
8 & 8 & 3.5 & 3.5 & 37 & 18.0 & 93.0 & 45 & 191.7 & 1155.6 \\
9 & 9 & 4.0 & 4.0 & 46 & 22.5 & 130.7 & 55 & 227.7 & 1312.2 \\
\hline
\end{tabular}
\caption{$R_{\Psi}^{*}$ for the treatment, interaction and quadratic models for the $D$- and $A$-optimality criteria. Results are reported for $s=1,\ldots,9$. } \label{tab:LargeSample}
\end{table}

\section{Simulation Study} \label{sec:sim}
In this section a simulation study is conducted to compare the finite sample efficiency of the ROAD and FOD. The simulation study considers the three different models listed in the preceding section. Note the ROAD design is a likelihood based method and it depends on the distribution of the errors. Two error distributions will be considered as illustrative examples.

\emph{STR errors}: The first error distribution considered is the Student t-distribution (STR) with parameter $v$. The density of the residuals is proportional to
$f(\varepsilon) \propto (1 + \varepsilon^{2}/v)^{-(1+v)/2}$. A STR with $v=1$ corresponds to the Cauchy distribution.
The statistical curvature for the STR model is given in \cite{Efro:Defi:1975} where it is shown that $\gamma \rightarrow \infty$ as $v\rightarrow 0$ and is a decreasing function of $v$. For a STR with $v=$ 1/2, 1 and 2 $\gamma^{2} \approx$ 5.6, 2.5 and 1.1, respectively.

\emph{GHR errors}: A second error distribution is Fisher's gamma hyperbola model. The gamma hyperbola density is $f(\varepsilon,a) \propto a^{v-1}e^{2a{\rm{Cosh}}(\varepsilon)}$,
where $\rm{Cosh}(\varepsilon) = (e^{\varepsilon} +e^{-\varepsilon})/2$ and $a$ is an ancillary random variable given in the supplemental materials. In gamma hyperbola regression (GHR) $\gamma^{2} = (2v)^{-1}$; therefore, $\gamma\rightarrow\infty$ as $v\rightarrow 0$.

\subsection{Efficiency Results}
In this section the efficiency results of the simulation study are presented. Note, the results for $D$-optimality are reported in terms of $|J_{\boldsymbol{A}}|^{1/2}$. Recall the volume of the confidence ellipsoid is proportional to the square root of the determinant; therefore, $|J_{\boldsymbol{A}}|^{1/2}$ represents the efficiency on the scale of the volume of the confidence ellipsoid.

For each of the two optimality criteria the simulation study presents the $\Psi\mbox{-Eff}_{\rm{CI}}$. This measure describes the efficiency of the confidence regions / power of the ROAD relative to the FOD. Values greater than one indicate that the ROAD results in smaller confidence regions or greater power, depending on the optimality criteria. 

In Section \ref{sec:Eff} $\Psi\mbox{-Eff}_{\rm{CMSE}}$ was defined to describe the efficiency of the ROAD relative to the expected conditional MSE. Unfortunately, there is no known closed form expression available for the conditional MSE in the examples considered. Instead $\Psi\mbox{-Eff}_{\rm{UMSE}}$ is reported in the simulation study. Note, the technical results given in Section \ref{sec:Eff} do not hold for this measure; despite this fact $\Psi\mbox{-Eff}_{\rm{UMSE}}$ was still greater than one in nearly every case considered, which might indicate that the theoretical results extend, in some fashion, to this measure.

\begin{figure}
\setlength{\tempwidth}{.48\linewidth}
\settoheight{\tempheight}{\includegraphics[width=\tempwidth]{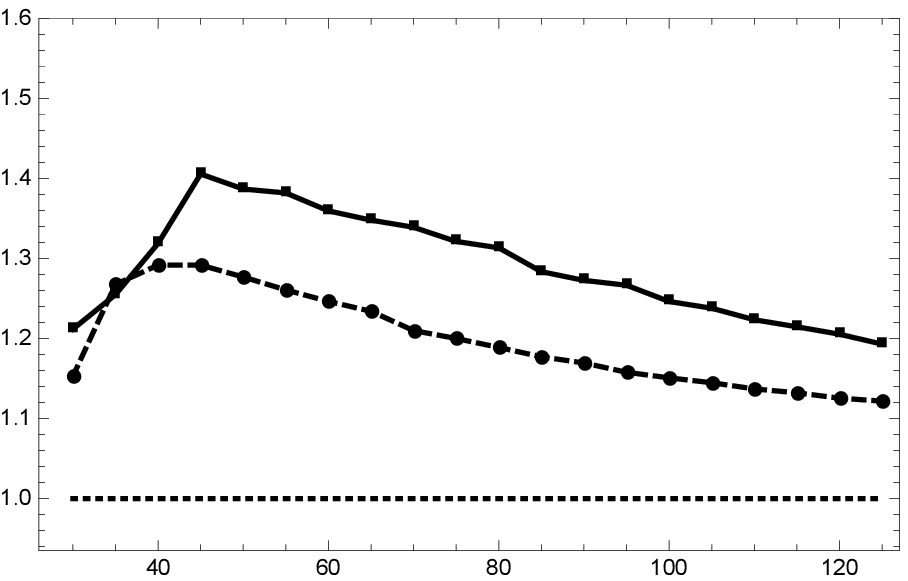}}
\centering
\hspace{\baselineskip}\\
\rowname{{\normalfont\scriptsize Treatment $s=4$}}
\begin{subfigure}[b]{\tempwidth}
       \centering
       \includegraphics[width=\tempwidth]{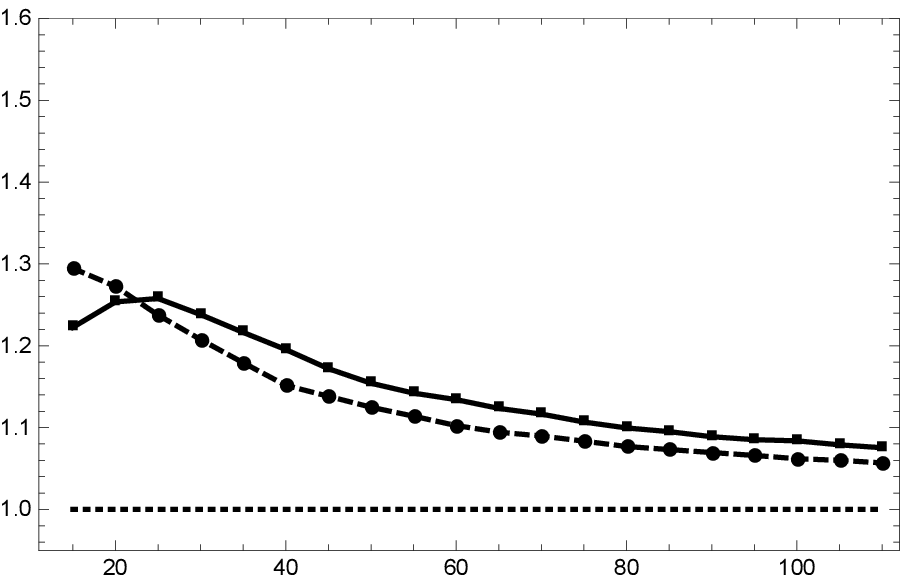}
\end{subfigure}
\begin{subfigure}[b]{\tempwidth}
       \centering
       \includegraphics[width=\tempwidth]{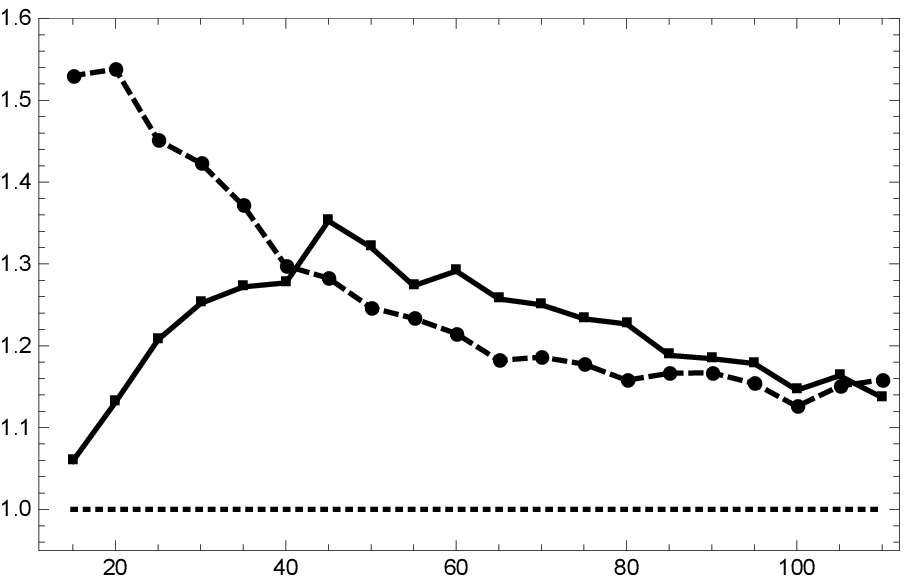}
\end{subfigure} \\
\rowname{{\normalfont\scriptsize Treatment $s=6$}}
\begin{subfigure}[b]{\tempwidth}
       \centering
       \includegraphics[width=\tempwidth]{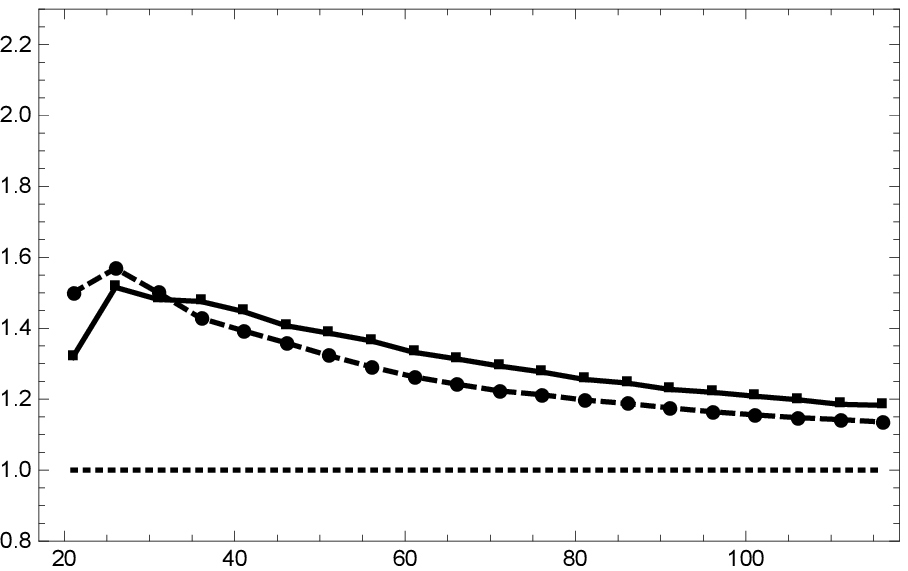}
\end{subfigure}
\begin{subfigure}[b]{\tempwidth}
       \centering
       \includegraphics[width=\tempwidth]{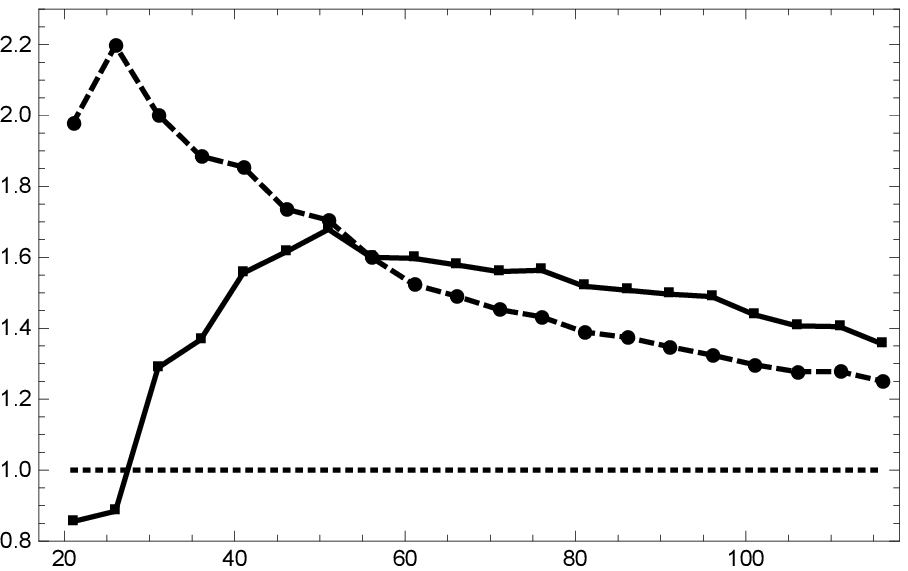}
\end{subfigure}\\
\settoheight{\tempheight}{\includegraphics[width=\tempwidth]{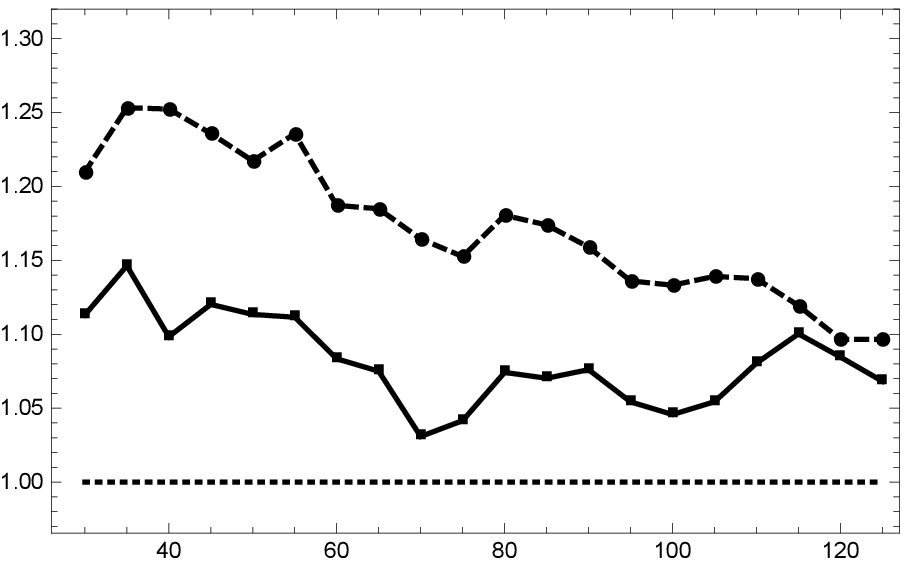}}
\rowname{{\normalfont\scriptsize Interaction $s=3$}}
\begin{subfigure}[b]{\tempwidth}
       \centering
       \includegraphics[width=\tempwidth]{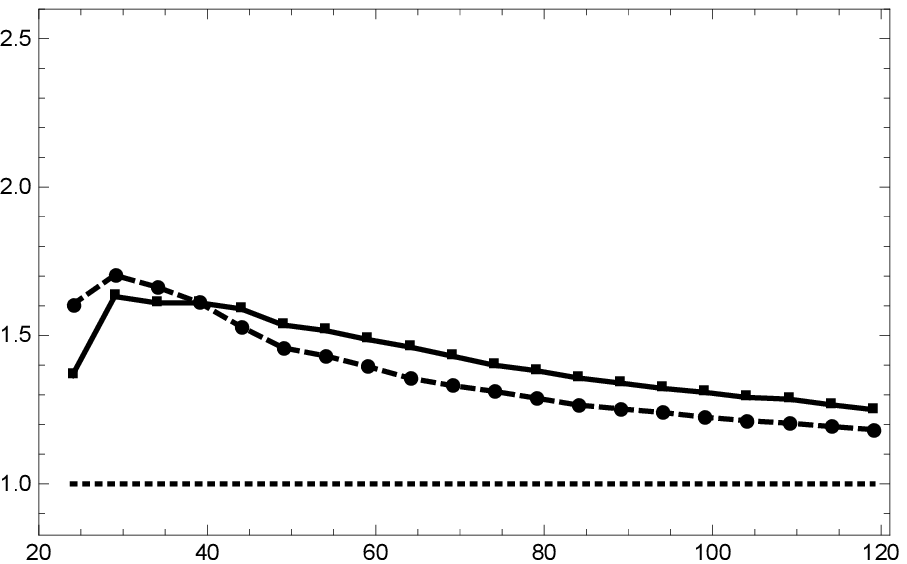}
\end{subfigure}
\begin{subfigure}[b]{\tempwidth}
       \centering
       \includegraphics[width=\tempwidth]{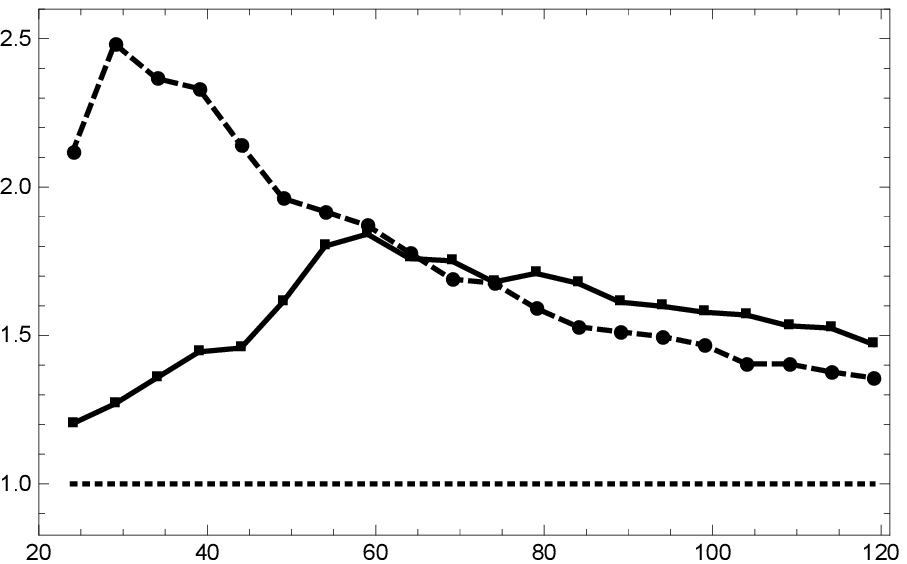}
\end{subfigure} \\
\rowname{{\normalfont\scriptsize Quadratic $s=2$}}
\begin{subfigure}[b]{\tempwidth}
       \centering
       \includegraphics[width=\tempwidth]{OD_qs2.eps}
       %\caption{${\rm{REff}}_{\rm{Inf}}$}
\end{subfigure}
\begin{subfigure}[b]{\tempwidth}
       \centering
       \includegraphics[width=\tempwidth]{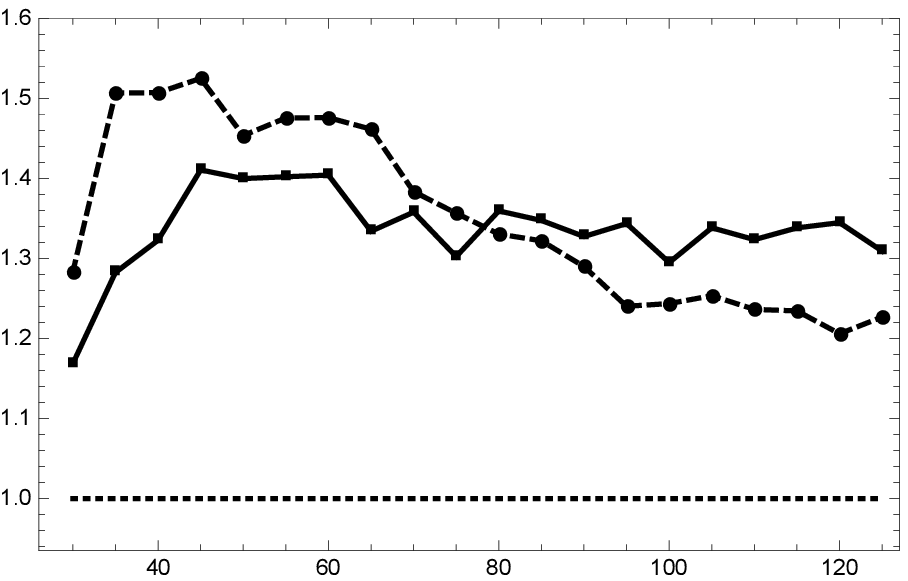}
       %\caption{$\Psi\rm{-Eff}_{\rm{CMSE}}$}
\end{subfigure}
\columnname{\normalfont\scriptsize (a)  $D\mbox{-Eff}_{\rm{CI}}$}
\columnname{\normalfont\scriptsize (b)  $D\mbox{-Eff}_{\rm{UMSE}}$}
\caption{Simulation of the efficiencies, from 10,000 iterations, for (a) $D\mbox{-Eff}_{\rm{CI}}$ and (b) $D\mbox{-Eff}_{\rm{UMSE}}$. Each sub-figure presents efficiencies for STR errors (solid curve), with $v=1$, and GHR errors, with $v=1/4$ (dashed curve). The rows, from top to bottom, correspond to the treatment model with $s=4$, the treatment model with $s=6$, the interaction model with $s=3$ and the quadratic model with $s=2$. The sample size range for each simulation is $n = kd+(3,103)$. For every example $\boldsymbol{\beta}=\boldsymbol{1}_{p}$ and $k=3$. In each figure a horizontal line at one is plotted for reference. A solid or dashed curve is greater than one indicates that the ROAD is more efficient than the FOD.}\label{fig:DOpt}
\end{figure}

\begin{figure}
\setlength{\tempwidth}{.48\linewidth}
\settoheight{\tempheight}{\includegraphics[width=\tempwidth]{OD_qs2.eps}}
\centering
\hspace{\baselineskip}\\
\rowname{{\normalfont\scriptsize Treatment $s=4$}}
\begin{subfigure}[b]{\tempwidth}
       \centering
       \includegraphics[width=\tempwidth]{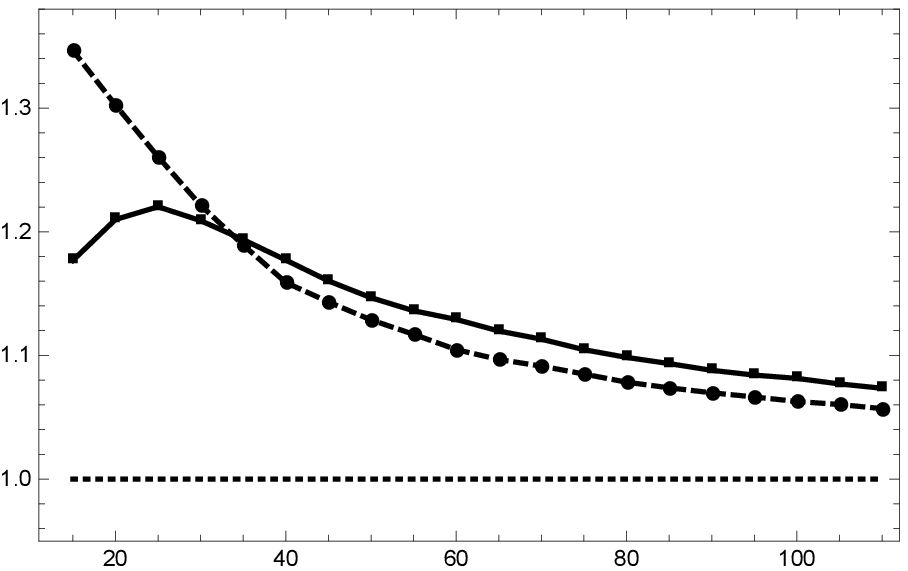}
\end{subfigure}
\begin{subfigure}[b]{\tempwidth}
       \centering
       \includegraphics[width=\tempwidth]{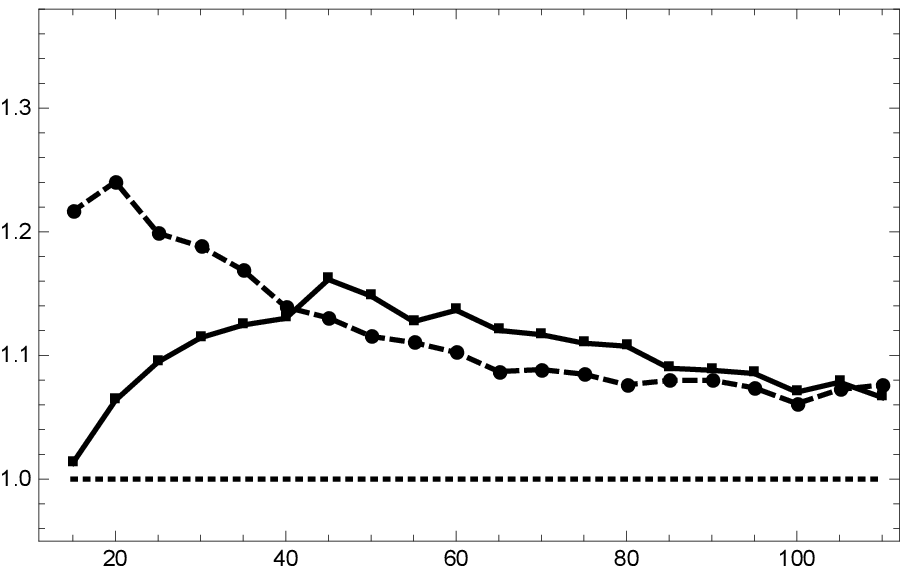}
\end{subfigure} \\
\rowname{{\normalfont\scriptsize Treatment $s=6$}}
\begin{subfigure}[b]{\tempwidth}
       \centering
       \includegraphics[width=\tempwidth]{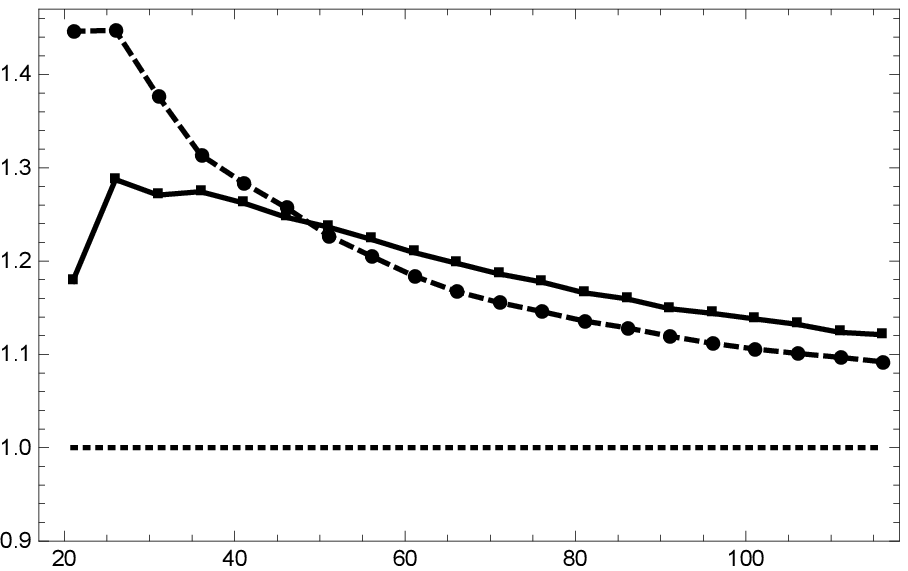}
\end{subfigure}
\begin{subfigure}[b]{\tempwidth}
       \centering
       \includegraphics[width=\tempwidth]{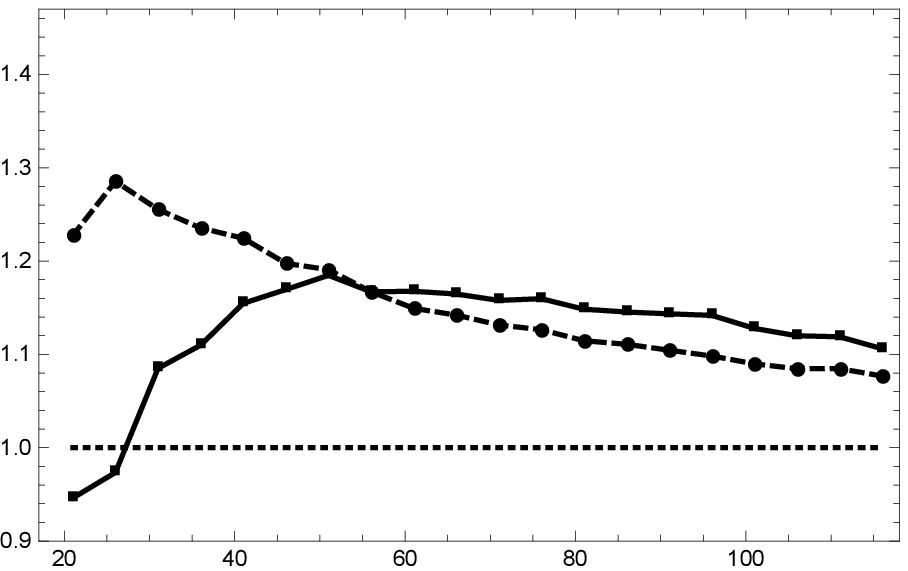}
\end{subfigure}\\
\settoheight{\tempheight}{\includegraphics[width=\tempwidth]{OD_qs2.eps}}
\rowname{{\normalfont\scriptsize Interaction $s=3$}}
\begin{subfigure}[b]{\tempwidth}
       \centering
       \includegraphics[width=\tempwidth]{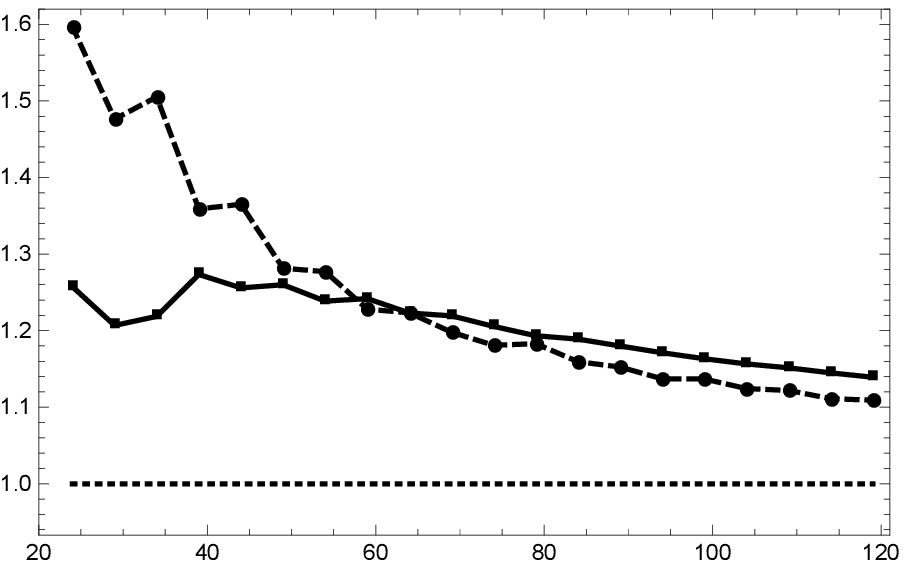}
\end{subfigure}
\begin{subfigure}[b]{\tempwidth}
       \centering
       \includegraphics[width=\tempwidth]{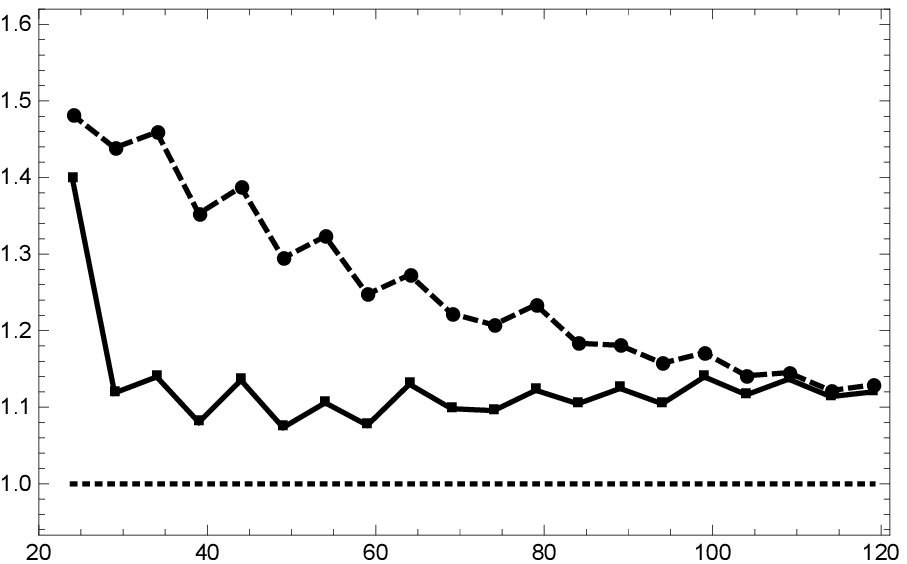}
\end{subfigure} \\
\rowname{{\normalfont\scriptsize Quadratic $s=2$}}
\begin{subfigure}[b]{\tempwidth}
       \centering
       \includegraphics[width=\tempwidth]{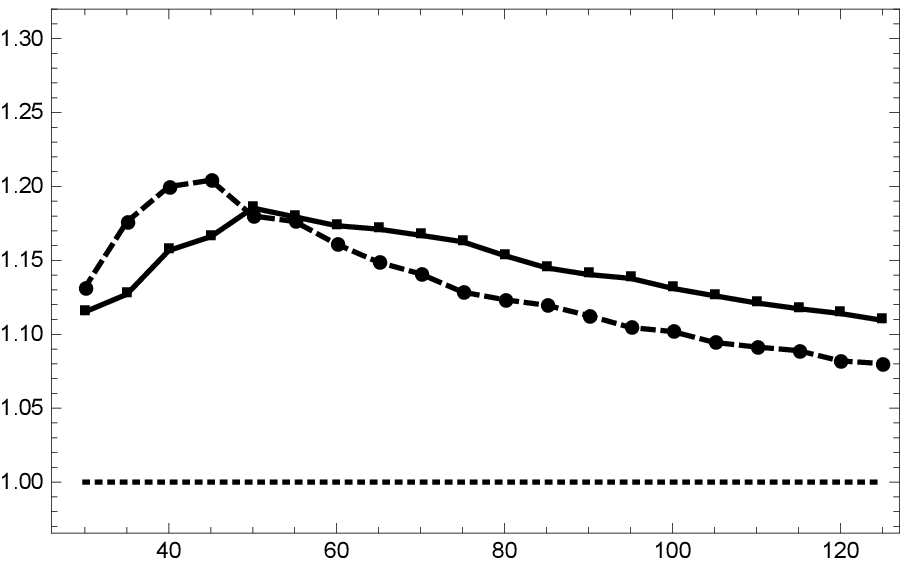}
       %\caption{${\rm{REff}}_{\rm{Inf}}$}
\end{subfigure}
\begin{subfigure}[b]{\tempwidth}
       \centering
       \includegraphics[width=\tempwidth]{VA_qs2.eps}
       %\caption{$\Psi\rm{-Eff}_{\rm{CMSE}}$}
\end{subfigure}
\columnname{\normalfont\scriptsize (a)  $A\mbox{-Eff}_{\rm{CI}}$}
\columnname{\normalfont\scriptsize (b)  $A\mbox{-Eff}_{\rm{UMSE}}$}
\caption{Simulation of the efficiencies, from 10,000 iterations, for (a) $A\mbox{-Eff}_{\rm{CI}}$ and (b) $A\mbox{-Eff}_{\rm{UMSE}}$. Each sub-figure presents efficiencies for STR errors (solid curve), with $v=1$, and GHR errors, with $v=1/4$ (dashed curve). The rows, from top to bottom, correspond to the treatment model with $s=4$, the treatment model with $s=6$, the interaction model with $s=3$ and the quadratic model with $s=2$. The sample size range for each simulation is $n = kd+(3,103)$. For every example $\boldsymbol{\beta}=\boldsymbol{1}_{p}$ and $k=3$. In each figure a horizontal line at one is plotted for reference. A solid or dashed curve is greater than one indicates that the ROAD is more efficient than the FOD.}\label{fig:AOpt}
\end{figure}

Figures \ref{fig:DOpt} and \ref{fig:AOpt} present the efficiencies, from 10,000 iterations, for $D$- and $A$-optimality, respectively, for $\Psi\mbox{-Eff}_{\rm{CI}}$ [column (a)] and $\Psi\mbox{-Eff}_{\rm{UMSE}}$ [column (b)]. In each sub-figure the solid and dashed curves corresponds to STR errors, with $v=1$, and the GHR errors, with $v=1/4$, respectively. The rows of Figures \ref{fig:DOpt} and \ref{fig:AOpt}, from top to bottom, present the results for the treatment model with $s=4$, the treatment model with $s=6$, the interaction model with $s=3$ and the quadratic model with $s=2$. For every example $\boldsymbol{\beta}=\boldsymbol{1}_{p}$ and $k=3$. Note, the sample size range differs for each example. The sample size range for each simulation is $n = kd+(3,103)$. For example, the sample size ranges from $n=15$ to 115 for treatment model with $s=4$. In each figure a horizontal line at one is plotted for reference. A solid or dashed curve greater than one indicates that the ROAD is more efficient than the FOD. 

First consider $\Psi\mbox{-Eff}_{\rm{CI}}$ in column (a) of each figure. Corollary \ref{cor:opt} proved that for large $n$ the efficiency of the ROAD is greater than the FOD with respect to this measure. The results of the simulation study show that for finite sample sizes this is true for every optimality criteria, error distribution, mean function and sample size considered. The difference can be substantial with the largest benefit when the sample size is smaller. For example, for the interaction model with $s=3$ and $n=29$ the $D$-efficiency of the ROAD relative to the FOD is approximately 1.6 and 1.7 for STR and GHR errors, respectively. This translates to a volume of the confidence ellipsoid for the FOD being approximately 1.7, or 1.6, times larger than the corresponding volume for the ROAD depending on the error distribution. This benefit persists to moderate sample sizes; for the same example with $n=124$ the efficiency of the ROAD relative to the FOD is approximately 1.25 and 1.18 for STR and GHR errors, respectively. Note, the relationships given in Theorems \ref{thm:MSE} and \ref{cor:efro:hink} between $J_{\boldsymbol{A}}$ and conditional MSE imply that $\Psi\mbox{-Eff}_{\rm{CMSE}}$ might have a similar improvement to $\Psi\mbox{-Eff}_{\rm{CI}}$. However, as stated, there is no known closed form expression available for the conditional MSE and the ROAD and FOD have not been directly compared with respect to  $\Psi\mbox{-Eff}_{\rm{CMSE}}$.

The results of the simulation for $\Psi\mbox{-Eff}_{\rm{CI}}$ also demonstrate the impact of statistical curvature; $\gamma^{2}$ is equal to 2.5 and 2 for STR and GHR errors, respectively. This means that for large $n$ it is expected that the efficiency of the ROAD relative to the FOD is greater for STR errors than GHR errors. Note for small sample sizes this behavior is not always observed. In fact, in seven out of the eight examples the STR has lower relative efficiency for the first sample size presented. However, as the sample size increases the relative efficiency of the ROAD is greater for STR errors as expected.  

Next consider the results for $\Psi\mbox{-Eff}_{\rm{UMSE}}$ in column (b) of Figures \ref{fig:DOpt} and \ref{fig:AOpt}. As previously stated, there are no theoretical results given that establishes that the ROAD is more efficient that the FOD with respect to this measure. However, intuitively, the result of Theorem \ref{thm:MSEopt}, that $\Psi\mbox{-Eff}_{\rm{CMSE}}>1$, may translate to a benefit in unconditional MSE measured by $\Psi\mbox{-Eff}_{\rm{UMSE}}$. This intuition is proven correct in column (b) of each figure. In all but four cases considered the $\Psi\mbox{-Eff}_{\rm{UMSE}}>1$, indicating that the ROAD is more efficient than the FOD with respect to unconditional MSE. Again, the benefit can be substantial. For the interaction model with $s=3$ and $n=29$ the efficiency of the ROAD relative to the FOD is approximately 1.3 and 2.5 for STR and GHR errors, respectively. This benefit persists to moderate sample sizes; for the same example with $n=124$ the ROAD relative to the FOD is approximately 1.5 and 1.4 for STR and GHR errors, respectively. 

The four exceptions, where the FOD was more efficient than the ROAD, occurred for the two smallest sample sizes for each criteria in treatment model with $s=6$ and STR errors. This highlights a pattern that can be observed in many of the other examples. After initialization a burn in period of several additional samples may be required before the full benefit of the adaptive procedure can be realized. This is something to be aware of the experiment only has a very small number of additional samples available post-initialization.


\subsection{Power Results}

A $\boldsymbol{c}$-optimal design minimizes the unconditional MSE of the linear combination of the MLE, $\boldsymbol{c}^{T}\boldsymbol{\hat{\beta}}$, where $\boldsymbol{c}$ is a known vector of constants. In $\boldsymbol{c}$-optimal design $\Psi(M) =\boldsymbol{c}(M) = (\boldsymbol{c}^{T}M^{-1}\boldsymbol{c})^{-1}$. For the treatment example the $A$-optimal design is equivalent to a $\boldsymbol{c}$-optimal design with $\boldsymbol{c} = \boldsymbol{1}_{p}$. This is a result of the diagonal nature of the information matrices. In this section the power of the ROAD relative to the FOD will be evaluated with respect to this linear combination.

Under the null hypothesis, that ${\rm{E}}[\boldsymbol{c}^{T}\hat{\boldsymbol{\beta}}] = C_{0}$,
\begin{align} \label{eq:chi2}
\boldsymbol{c}(J_{\mathcal{A}})(\boldsymbol{c}^{T}\hat{\boldsymbol{\beta}} - C_{0})^{2} | \mathcal{A} \rightarrow \chi_{1}^{2}
\end{align}
in distribution as $n\rightarrow\infty$. In the presence of an alternative hypothesis, say ${\rm{E}}[\boldsymbol{c}^{T}\hat{\boldsymbol{\beta}}] = C_{1}$, the power of a $\chi^{2}$-test is calculated as
$\mbox{Power} = 1 - P\{\chi_{1}^{2}(\lambda)\ge \chi_{1}^{2{(1-\alpha)}}\}$,
where $\chi_{1}^{2}(\lambda)$ is non-central $\chi^{2}$-distribution with one degree of freedom and non-centrality parameter $\lambda$. In the current context the non-centrality parameter is
$\lambda_{\boldsymbol{A}} = n\delta^{2}\boldsymbol{c}(J_{\boldsymbol{A}})$,
where $\delta = C_{1} - C_{0}$. Power is a non-decreasing function of the non-centrality parameter. A ROAD with the $\boldsymbol{c}$-criterion will maximize the power of this test. The power analysis presented in this section uses $\boldsymbol{c} = \boldsymbol{1}_{p}$, $\boldsymbol{\beta} = \boldsymbol{1}_{p}/2$, $C_{0} = 0$ and a variety of sample sizes. 

\begin{figure}
\setlength{\tempwidth}{.48\linewidth}
\settoheight{\tempheight}{\includegraphics[width=\tempwidth]{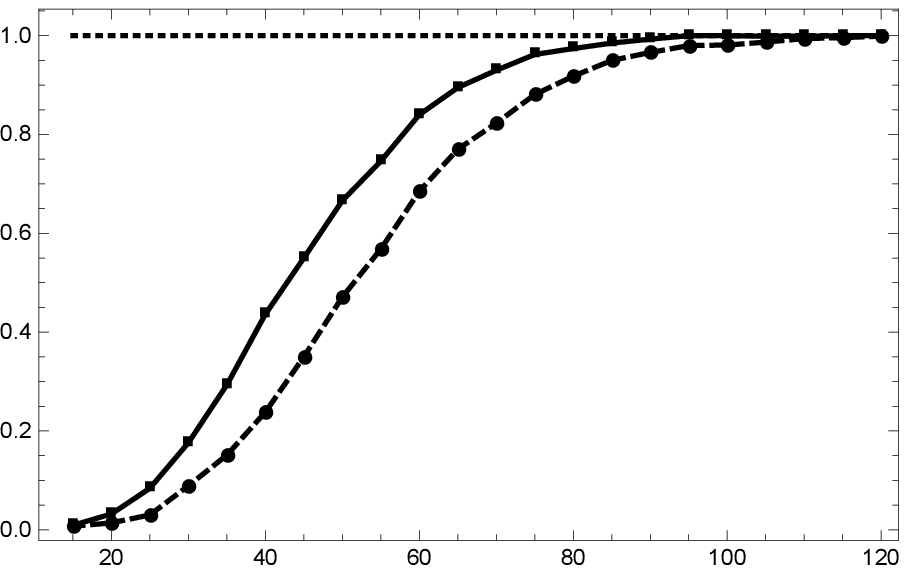}}
\centering
\hspace{\baselineskip}\\
\rowname{{\normalfont\scriptsize Treatment $s=4$}}
\begin{subfigure}[b]{\tempwidth}
       \centering
       \includegraphics[width=\tempwidth]{PowSTR_txt4.eps}
\end{subfigure}
\begin{subfigure}[b]{\tempwidth}
       \centering
       \includegraphics[width=\tempwidth]{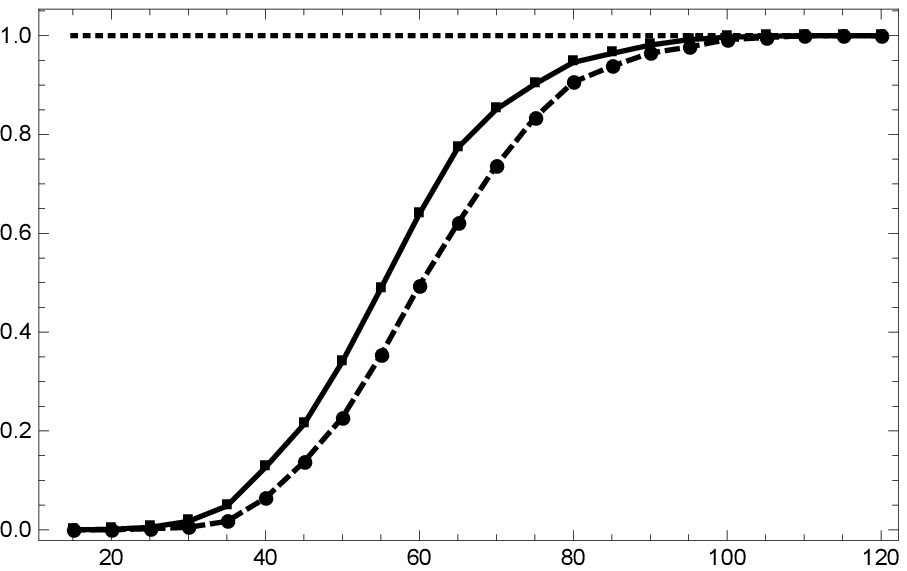}
\end{subfigure} \\
\rowname{{\normalfont\scriptsize Treatment $s=6$}}
\begin{subfigure}[b]{\tempwidth}
       \centering
       \includegraphics[width=\tempwidth]{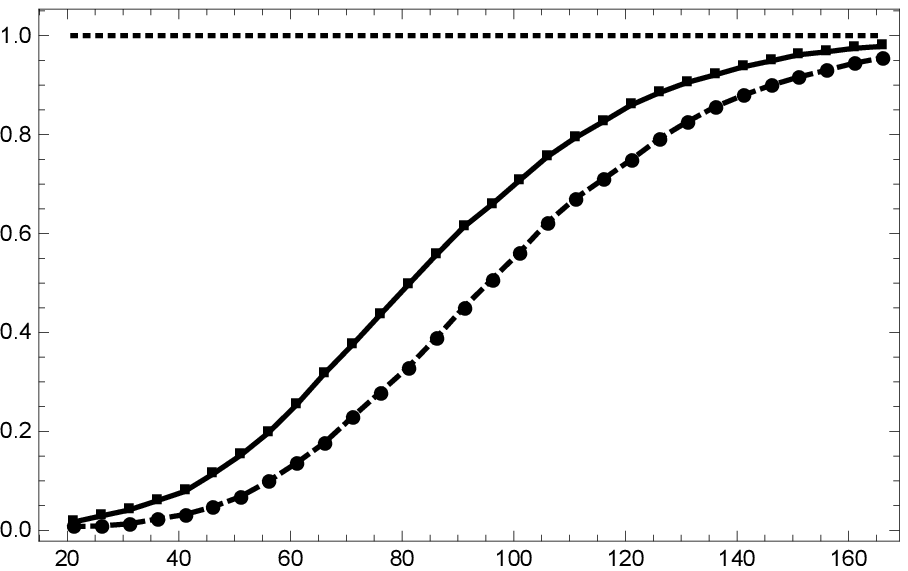}
\end{subfigure}
\begin{subfigure}[b]{\tempwidth}
       \centering
       \includegraphics[width=\tempwidth]{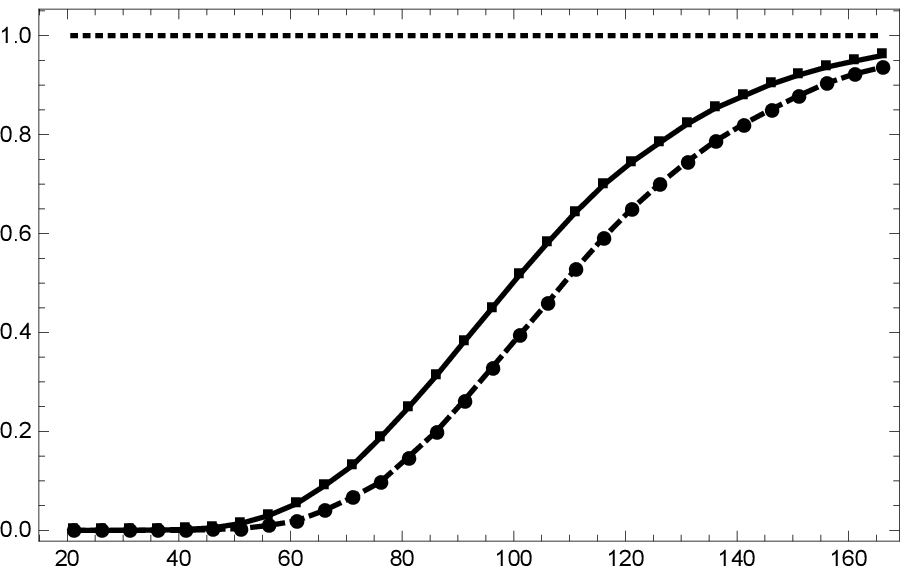}
\end{subfigure}\\
\columnname{\normalfont\scriptsize (a)  Power STR}
\columnname{\normalfont\scriptsize (b)  Power GHR}
\caption{Simulation power, from 10,000 iterations, of the $\chi^{2}$-test for the ROAD (solid line) and the FOD (dashed line). The treatment example with $p=4$ and $p=6$ are presented in the top and bottom rows, respectively, for (a) STR errors and (b) GHR errors. Values of $\boldsymbol{c} = \boldsymbol{1}_{p}$, $\boldsymbol{\beta} = \boldsymbol{1}_{p}/2$, $C_{0} = 0$ and $k=3$ were used for each example.The greater the power the better. In each figure horizontal line at one is plotted for reference.}\label{fig:cPow}
\end{figure}

Figure \ref{fig:cPow} presents the power of the $\chi^{2}$-test, from 10,000 iterations, for the ROAD (solid line) and the FOD (dashed line). The treatment example with $p=4$ and $p=6$ are shown in the top and bottom rows, respectively. The STR and GHR power functions are in columns (a) and (b), respectively. The ROAD was uniformly more powerful than the FOD in all cases considered. For both $p=4$ and 6 the improvement in power is significant. For example, it is often desired to select the minimum sample size that attains a nominal power of 0.8 while maintaining type I error rate of $\alpha=0.05$. For STR errors with $p=6$ a ROAD with $n=112$ is the minimum sample size that attains this nominal power. For the same example the FOD required $n=128$, an additional 16 observations, to achieve the same nominal power. 

\section{Discussion}

In this work the linear regression observed information adaptive design (ROAD) was proposed. The primary objective of this design is to  optimize the observed Fisher information subject to a concave optimality criterion, $\Psi$. The method was contrasted against the fixed optimal design (FOD), which maximizes $\Psi$ evaluated at the expected Fisher information. The focus of the comparison was on optimizing conditional inference and MSE. Optimal inference was defined as maximizing $\Psi$ evaluated at the observed Fisher information. A design is optimal with respect to conditional MSE if it maximizes $\Psi$ evaluated at the inverse of the conditional MSE. In this work it is shown that for every fixed design, including the FOD, there exists a ROAD that is more efficient with respect to both conditional inference and  MSE, at the limit.

Historically, adaptive designs have been constrained to experiments where the design objective depends on characteristics of the underlying population, e.g. the model parameters. Such population based adaptive designs (PADs) update information about the population using the data available from the preceding runs to determine the design of the next run, see Section \ref{sec:Intro}. Since, the primary objective of PADs is to update information about the underlying population it stands to reason that if this information were known in advance then a fixed design could be determined that is as efficient, if not more so, than the relevant PAD. Observed information adaptive designs (OADs), e.g. the ROAD in this work and the LOAD and MOAD in \cite{Lane:Adap:2019}, represent a new paradigm in adaptive design where the primary objective is to accommodate new information about the \emph{observed sample} and not new information about the \emph{underlying population} into the design. The OAD framework allows adaptive designs to be applied to experiments that previously would not benefit from PADs. 

In addition to the theoretical results a simulation study was conducted to compare the ROAD and FOD. The ROAD was more efficient than the FOD with respect to unconditional MSE, inference and power in nearly every case considered. This suggests that the large sample benefits found in the theoretical results are also present in finite sample sizes. The primary conclusion is that using observed Fisher information in an adaptive design can significantly improve the precision of the MLE. 

Focusing on a linear model with additive errors made the theoretical results tractable. A more general model was considered in \cite{Lane:Adap:2019}, where it was shown, primarily with heuristics and a simulation study, that observed information adaptive designs can reduce the MSE of the parameter estimates. Based on the agreement between the simulation study in Section \ref{sec:sim} and that of \cite{Lane:Adap:2019} it might be conjectured that the theoretical results extend to more general models. Extending the theoretical findings in Section \ref{sec:Cond} is far from trivial.

The primary weakness of this method is that it is likelihood based. Meaning that it is required that the error distribution is known. Methods that are free of distributional assumptions, e.g. the least squares estimates, are still asymptotically optimized by the FOD. The sequential nature of the data collection allows this weakness to be remedied. In step 1 of the ROAD algorithm $kd$ observations are collected. This step is used to initialize the design and ensure the reliability of the observed Fisher information before beginning the adaptive phase of the design. However, the data collected from step 1 could also be used to plot the residuals against a set of candidate error distributions. The likelihood of the error distribution with the best fit could then be used to calculate the observed Fisher information for the adaptive phase. 

\section{Technical Details} \label{sec:Tech}

In this section the supporting arguments for the main theoretical results are presented.

\subsection{Technical Conditions}
Denote the $k$th derivative of the log likelihood of the $ij$th observation as
$l_{0}^{(\cdot k)}(\varepsilon_{ij}) = (\partial^{k}/\partial \eta_{i}^{k}) \log f_{0}(y_{ij} - \eta_{i})$. 
\begin{condition}\label{C1}
%\begin{minipage}[t]{\linewidth}
Conditions on the distribution of responses: (1) $\boldsymbol{\beta}\in B$, where $B$ is an open subset of $\mathbb{R}^{p}$; (2) $\boldsymbol{\varepsilon}$ is a vector of independent and identically distributed random variables such that $f_{\eta}(y) = f_{0}(\varepsilon)$; (3) $l_{0}^{(\cdot k)}(\varepsilon_{ij})$ exists, E$[|l_{0}^{(\cdot k)}(\varepsilon_{ij})|]<\infty$ and the derivative $\partial^{k}/\partial\eta_{i}^{k}$ can be exchanged with the expectation for $k=1,\ldots,4$; (4) E$[\ddot{l}_{0}(\varepsilon_{ij})]<0$; and (5) $l_{0}^{(\cdot 5)}(\varepsilon_{ij})$ is bounded in probability.
\end{condition}

\begin{condition}\label{C2}
Conditions on the design: (1)
$\Psi(\cdot)$ is a non-negative concave positive-homogeneous function with degree $1$ and continuous and bounded first and second derivatives; (2) a design $\xi_{\Psi}^{*}$ is known such that $\Psi[M(\xi_{\Psi}^{*})]= \Psi_{\Delta}^{*}[1+O(n^{-1})]$ and has a finite number of support points; (3) $M_{\Psi}^{*}$ and $H_{\Psi}^{*}$ are positive definite and $\Psi^{*}>0$; and (4)
$\mathcal{X}$ is a compact subspace of $\mathbb{R}^{s}$. 
%\end{minipage}
\end{condition}

\subsection{Proof of Theorem \ref{thm:MSE}}
In this theorem the design is assumed to be fixed. The necessary conditions for the theorem to hold are Condition \ref{C1}, ${\rm{E}}[(\hat{\eta}_{i} - \eta_{i})^{3}]<\infty$ and that the allocation weights of the fixed design $\xi_{n}$ are positive constants as $n\rightarrow\infty$. Note under the stated conditions $\sqrt{n_{i}}(\hat{\eta}_{i} - \eta_{i})$ has an asymptotically standard normal distribution.

From \eqref{eq:trans_dist} we can write the likelihood as
$L(\boldsymbol{\beta}|\boldsymbol{y}) \propto \prod_{i=1}^{d} g_{\boldsymbol{a_{i}}}[\hat{\eta}_{i} - \boldsymbol{\beta}^{T}f_{x}(x_{i})]$. By the factorization theorem it is follows that $(\boldsymbol{\hat{\eta}},\mathcal{A})$ is a sufficient statistic. Since, $\boldsymbol{\hat{\beta}}$ given $\mathcal{A}$ is an injection of $\boldsymbol{\hat{\eta}}$ it follows that $(\boldsymbol{\hat{\beta}},\mathcal{A})$ is a sufficient statistic.

Note that $\dot{l}_{\eta_{i}}(y_{ij}) = l_{\eta_{i}}^{(\cdot 1)}(y_{ij})$, $\ddot{l}_{\eta_{i}}(y_{ij}) = l_{\eta_{i}}^{(\cdot 2)}(y_{ij})$ and $l_{\eta_{i}}^{(\cdot k)}(y_{ij}) = l_{0}^{(\cdot k)}(\varepsilon_{ij})$. Further, let $l_{n_{i}}^{(\cdot k)}(\boldsymbol{\varepsilon_{i}}) = \sum_{j=1}^{n_{i}} l_{0}^{(\cdot k)}(\varepsilon_{ij})$ and $l_{n}^{(\cdot k)}(\boldsymbol{\varepsilon}) = \sum_{i=1}^{d} l_{n_{i}}^{(\cdot k)}(\boldsymbol{\varepsilon_{i}})$. Remark that the $k$th derivative of the log likelihood for the $i$th design point evaluated the MLE, $l_{n_{i}}^{(\cdot k)}(0) = l_{\boldsymbol{a_{i}}}^{(\cdot k)}$, is a function of the data through $\boldsymbol{a_{i}}$ alone. 

The score function for $\boldsymbol{\beta}$ can be written as $U(\boldsymbol{\beta}) = \sum_{i=1} f_{x}^{T} \dot{l}_{n_{i}}(e_{i})$,
where $e_{i} = \hat{\eta}_{i} - \eta_{i}$. A Taylor expansion of $ U(\boldsymbol{\beta})$ for $e_{i}$ around 0, for $i=1,\ldots,d$, yields
\begin{align}
    U(\boldsymbol{\beta}) 
    &= \sum_{i=1}^{d} f_{x}(x_{i})^{T} \left[-e_{i}\boldsymbol{i}_{\boldsymbol{a_{i}}} + \frac{1}{2}e_{i}^{2}l_{\boldsymbol{a_{i}}}^{(\cdot 3)} + \frac{1}{6}e_{i}^{3}l_{\boldsymbol{a_{i}}}^{(\cdot 4)}\right] + o_{p}(n^{-1/2}) \\
    &= -F^{T} I_{\boldsymbol{A}} \boldsymbol{e} + \frac{1}{2}F^{T}K_{\boldsymbol{A}} diag(\boldsymbol{e})\boldsymbol{e} + \frac{1}{6}F^{T}  L_{\boldsymbol{A}} diag(\boldsymbol{e}^{T}\boldsymbol{e})\boldsymbol{e}+ o_{p}(n^{-1/2}),
\end{align}
where $K_{\boldsymbol{A}} = diag(l_{\boldsymbol{a_{1}}}^{(\cdot 3)},\ldots,l_{\boldsymbol{a_{d}}}^{(\cdot 3)})^{T}$, $L_{\boldsymbol{A}} = diag(l_{\boldsymbol{a_{1}}}^{(\cdot 4)},\ldots,l_{\boldsymbol{a_{d}}}^{(\cdot 4)})^{T}$ and $\boldsymbol{e} = \boldsymbol{\hat{\eta}} - F \boldsymbol{\beta}$. 

Note $U(\boldsymbol{\hat{\beta}})$ has the same form as $U(\boldsymbol{\beta})$ with $\boldsymbol{e}$ replaced with $\boldsymbol{\hat{e}}=\boldsymbol{\hat{\eta}} - F \boldsymbol{\hat{\beta}}$. Then recalling that $J_{\boldsymbol{A}} = F^{T} I_{\boldsymbol{A}} F$ and recognizing that $U(\boldsymbol{\hat{\beta}}) = 0$ we get after some algebra that
\begin{align}
    \boldsymbol{\hat{\beta}} -  \boldsymbol{\beta} = J_{\boldsymbol{A}}^{-1}F^{T} \left[I_{\boldsymbol{A}}\boldsymbol{e} - \frac{1}{2} K_{\boldsymbol{A}} diag(\boldsymbol{\hat{e}})\boldsymbol{\hat{e}} - \frac{1}{6}L_{\boldsymbol{A}} diag(\boldsymbol{\hat{e}}^{T}\boldsymbol{\hat{e}})\boldsymbol{\hat{e}}\right] + o_{p}(n^{-3/2}).
\end{align}
By a substitution the residuals can be written as as
\begin{align}
    \boldsymbol{\hat{e}} = \boldsymbol{\hat{\eta}} - F \boldsymbol{\hat{\beta}} = (I_{d} - P_{\boldsymbol{A}})\boldsymbol{e} - \frac{1}{2}FJ_{\boldsymbol{A}}^{-1} F^{T} K_{\boldsymbol{A}} diag(\boldsymbol{\hat{e}})\boldsymbol{\hat{e}} + O_{p}(n^{-3/2}), \label{eq:resid}
\end{align}
where $P_{\boldsymbol{A}} = FJ_{\boldsymbol{A}}^{-1} F^{T} I_{\boldsymbol{A}}$ is a $d\times d$ projection matrix with rank $p$. Using \eqref{eq:resid} we can now write
\begin{align} \label{eq:delta}
        \boldsymbol{\hat{\beta}} - \boldsymbol{\beta} &= J_{\boldsymbol{A}}^{-1}F^{T} \left[I_{\boldsymbol{A}}\boldsymbol{e} - \frac{1}{2}K_{\boldsymbol{A}} diag[(I_{d} - P_{\boldsymbol{A}})\boldsymbol{e}](I_{d} - P_{\boldsymbol{A}})\boldsymbol{e}\right] + O_{p}(n^{-3/2}).
\end{align}

The following lemma considers the first 3 conditional moments of $e_{i}$.
\begin{lemma} \label{lem:efro:hink} Under the present conditions the following hold
\begin{align}
E[e_{i}|\boldsymbol{a_{i}}] &= -\frac{1}{2}l_{\boldsymbol{a_{i}}}^{(\cdot 3)} \{\boldsymbol{i}_{\boldsymbol{a_{i}}}\}^{-2} + o_{p}(n^{-1}), \quad
E[e_{i}^{2}|\boldsymbol{a_{i}}] = \{\boldsymbol{i}_{\boldsymbol{a_{i}}}\}^{-1} + o_{p}(n^{-1}), \\
E[e_{i}^{3}|\boldsymbol{a_{i}}] &= -\frac{15}{6}l_{\boldsymbol{a_{i}}}^{(\cdot 3)} \{\boldsymbol{i}_{\boldsymbol{a_{i}}}\}^{-3} + o_{p}(n^{-2}) \quad \mbox{and} \quad
\sqrt{n_{i}}(\boldsymbol{i}_{\boldsymbol{a_{i}}}/n_{i} - \mu) \rightarrow N(0,\gamma^{2})
\end{align}
for $i=1,\ldots,d$, where $\gamma$ is given by equation \eqref{eq:gamma}.
\end{lemma}
The first two conditional moments of $e_{i}$ and the asymptotic distribution of $\boldsymbol{i}_{\boldsymbol{a_{i}}}$ are proven in \cite{Efro:Hink:Asse:1978}. The proof for the 3rd conditional moments is obtained using very similar steps as those used in \cite{Efro:Hink:Asse:1978} and is omitted. Condition \ref{C1} and $E[(\hat{\eta}_{i} - \eta_{i})^{3}]$ are very similar to the conditions assumed in \cite{Efro:Hink:Asse:1978} and are required for the lemma.

From \eqref{eq:delta} the conditional MSE of $\hat{\boldsymbol{\beta}}$ can be written
\begin{align}
{\rm{MSE}}[\hat{\boldsymbol{\beta}}|\mathcal{A}] &= E[(\boldsymbol{\hat{\beta}} - \boldsymbol{\beta})(\boldsymbol{\hat{\beta}} - \boldsymbol{\beta})^{T}|\mathcal{A}] \\
&= J_{\mathcal{A}}^{-1}F^{T}E\left[I_{\mathcal{A}}\boldsymbol{e}\boldsymbol{e}^{T}I_{\mathcal{A}} - I_{\mathcal{A}}\boldsymbol{e}\boldsymbol{e}^{T}T_{\mathcal{A}}diag(\boldsymbol{e})K_{\boldsymbol{A}}|\mathcal{A} \right]FJ_{\mathcal{A}}^{-1} + O_{p}(n^{-2}),
\end{align}
where $T_{\boldsymbol{A}} = (I_{d} - P_{\boldsymbol{A}})^{T}diag[(I_{d} - P_{\boldsymbol{A}})]$. 

It can be shown that the $ij$th entry of $\boldsymbol{e}\boldsymbol{e}^{T}T_{\boldsymbol{A}}\boldsymbol{e}$ is $t_{ij}e_{i}e_{j}\sum_{k=1}^{d}t_{ik}e_{k}$, where $t_{ij}$ is the $ij$th entry of $T_{\boldsymbol{A}}$. From Lemma \ref{lem:efro:hink} and independence it can be shown that
$E[\boldsymbol{e}\boldsymbol{e}^{T}T_{\boldsymbol{A}}diag(\boldsymbol{e})|\mathcal{A}] = O_{p}(n^{-2})$. Which together with $E[\boldsymbol{e}\boldsymbol{e}^{T}|\mathcal{A}] = I_{\mathcal{A}}^{-1} + O_{p}(n^{-1})$ yields ${\rm{MSE}}[\hat{\boldsymbol{\beta}}|\mathcal{A}] = J_{\mathcal{A}}^{-1}[1+ O_{p}(n^{-1})]$ as stated.

The scaled difference between observed and expected Fisher information is $n^{-1}[J_{A} - \mathcal{F}(\xi)] = O_{p}(n^{-1/2})$ by lemma \ref{lem:efro:hink}. This implies that ${\rm{MSE}}[\hat{\boldsymbol{\beta}}|\mathcal{A}] = [\mathcal{F}(\xi)]^{-1}[1+ O_{p}(n^{-1})]$ as stated.

\subsection{Proof of Theorem \ref{thm:ancillary}}
For this theorem the necessary conditions are that a FOD, $\xi_{\Psi}^{*}$, is known and the errors are in the location family. The proof does not explicitly use the fact that $\xi_{\Psi}^{*}$ is known; however, as defined the ROAD only exists if such a design is known. 

We begin by establishing that in the location family the distribution of the residuals are unaffected by the ROAD procedure.
\begin{lemma} \label{lem:resid}
Under the present conditions, if $\boldsymbol{\check{\varepsilon}}$ is the vector of residuals from the ROAD procedure then
$f(\boldsymbol{\check{\varepsilon}}) = f(\boldsymbol{\varepsilon})$,
where $\boldsymbol{\varepsilon}$ is the vector of residuals obtained from any arbitrary fixed design.
\end{lemma}
Proof of lemma. For the ROAD procedure the data from the past observations is only used to determine the mean function of the present response. Therefore, conditional on the mean function the responses are independent of the past data. This independence property has been noted in adaptive designs \cite{Ford:Titt:Wu:infe:1985}. In the current setting this implies $\check{y}_{11}|\eta_{1},\ldots,\check{y}_{dn_{d}}|\eta_{d}$ are independent random variables. Therefore $(\check{y}_{11} - \eta_{1}|\eta_{1},\ldots,\check{y}_{dn_{d}} - \eta_{d}|\eta_{d}) = (\check{\varepsilon}_{11}|\eta_{1},\ldots,\check{\varepsilon}_{dn_{d}}|\eta_{d})$ are also independent random variables. However, $\check{\varepsilon}_{ij}$ are ancillary and thus the distribution of $\check{\varepsilon}_{ij}|\eta_{i}$ is the same as the distribution of $\check{\varepsilon}_{ij}$ which implies $\boldsymbol{\check{\varepsilon}}$ is a sequence of i.i.d random variables with the same distribution as if no adaptation had taken place. $\square$

Return to the proof of the theorem. First, the ROAD uses only the ancillary information available in the sample to determine the design for observation $j=2,\ldots,n$ and thus the design remains ancillary. 

After run $j$ the number of observations with design point $x_{i}^{*}$, denoted $N_{i}(j)$, is random. Denote the vector of responses observed at the $i$th design point from the first $j$ runs as $\boldsymbol{\check{y}_{i}}(j) = (\check{y}_{i1},\ldots,\check{y}_{i,N_{i}(j)})^{T}$. Further, define $\overline{\check{y}}_{i}(j)$ to be the mean of the responses corresponding to these observations. The ancillary configuration statistic can be represented as $\boldsymbol{\check{a}_{i}}(j) = [\check{y}_{1i} - \overline{\check{y}}_{i}(j), \ldots, \check{y}_{iN_{i}(j)} - \overline{\check{y}}_{i}(j)]^{T} = [\check{\varepsilon}_{1i}- \overline{\check{\varepsilon}}_{i}(j),\ldots,\check{\varepsilon}_{iN_{i}(j)} - \overline{\check{\varepsilon}}_{i}(j)]^{T}$ \cite{Fras:Anci:2004}. Recall, $\mathscr{A}(j)$ is a filtration and thus $\mathscr{A}(j-1)$ is $\mathscr{A}(j)$ measurable. Since $\boldsymbol{\check{a}_{i}}(j)$ can be written as a function of the errors it follows from Lemma \ref{lem:resid} that the $\boldsymbol{\check{a}_{i}}(j)$ remains ancillary following the ROAD. Since, $\mathscr{\check{A}}(j)$ is generated by $\mathcal{\check{A}}(k)$, $k=1,\ldots,j$ it too is ancillary. Further, per the definition of the ROAD the design point for the $j$th observation is $\mathscr{\check{A}}(j-1)$ measurable. In other words $\check{\boldsymbol{X}}$ is  $\mathscr{\check{A}}(n) = \mathscr{\check{A}}$ measurable. From Lemma \ref{lem:resid}
\begin{align}
\{\boldsymbol{\check{y}} - \boldsymbol{\check{\eta}}\}|\mathscr{\check{A}} \stackrel{d}{=} \check{\boldsymbol{\varepsilon}} \stackrel{d}{=} \boldsymbol{\varepsilon} \stackrel{d}{=} \{\boldsymbol{y} - \boldsymbol{\check{\eta}}\}|\{\boldsymbol{X} = \check{\boldsymbol{X}},\mathcal{A} = \mathcal{\check{A}}\}.
\end{align}
Subtracting $\boldsymbol{\check{\eta}}$ from both sides implies the stated result.

\subsection{Proof of Theorem \ref{thm:opt} }

This theorem requires Conditions \ref{C1}, \ref{C2} and $E[e_{i}^{2}]<\infty$. In what follows random variables with a check, eg. $\boldsymbol{\check{a}}_{i}$,  denote those obtained from a ROAD, random variables with a star, eg. $\boldsymbol{a^{*}}_{i}$, denote those obtained from a ROAD and random variables without a check or star, eg. $\boldsymbol{a}_{i}$, denote arbitrary random variables that could have been obtained from either a ROAD or FOD.

Let $\boldsymbol{\omega_{A}} = (\omega_{\boldsymbol{a}_{1}},\ldots\omega_{\boldsymbol{a}_{d}})^{T}$, where $\omega_{\boldsymbol{a}_{d}} = 1 - \sum_{i=1}^{d-1}\omega_{\boldsymbol{a}_{i}}$ and note that from the positive homogeneity assumption of $\Psi$ that
\begin{align} \label{eq:phObs}
\Psi\{J_{\boldsymbol{A}}\} = \Psi\{Q_{\boldsymbol{A}}M(\tau_{\boldsymbol{A}})/n\} = (Q_{\boldsymbol{A}}/n) \Psi\{M(\tau_{\boldsymbol{A}})\}.
\end{align}
A Taylor expansion of $\Psi\{M(\tau_{\boldsymbol{A}})\}$ with respect to $\boldsymbol{\omega}_{(d),\boldsymbol{A}}$ around $\boldsymbol{w}_{(d)}^{*}$, where the subscript $(d)$ indicates the $d$th entry of the vector is removed, yields
\begin{align} \label{eq:ObsExp}
\Psi\{M(\tau_{\boldsymbol{A}})\} &= \Psi^{*} + (\boldsymbol{\omega}_{(d),\boldsymbol{A}} - \boldsymbol{w}_{(d)}^{*})^{T}\nabla{\Psi}\{M(\xi^{*})\} \\
&\quad\quad  - (\boldsymbol{\omega}_{(d),\boldsymbol{A}}- \boldsymbol{w}_{(d)}^{*})^{T}H_{\Psi}^{*}  (\boldsymbol{\omega}_{(d),\boldsymbol{A}} - \boldsymbol{w}_{(d)}^{*})/2 \\
&\quad\quad + o_{p}\left[(\boldsymbol{\omega}_{(d),\boldsymbol{A}} - \boldsymbol{w}_{(d)}^{*})^{T}H_{\Psi}^{*}(\boldsymbol{\omega}_{(d),\boldsymbol{A}} - \boldsymbol{w}_{(d)}^{*})\right].
\end{align}
When $\xi^{*}$ is the continuous optimal design it the general equivalence theorem ensures that $\nabla{\Psi}\{M(\xi^{*})\}=0$. Is $\xi_{\Psi}^{*}$ is a design such that $\Psi[M(\xi_{\Psi}^{*})]= \Psi_{\Delta}^{*}[1+O(n^{-1})]$ then $\nabla{\Psi}\{M(\xi^{*})\}=O_{p}(n^{-1})$ [see supplemental materials]. Therefore, \eqref{eq:ObsExp} can be written, dropping the $o_{p}$ term, as
\begin{align} \label{eq:EffOAD}
    \Psi\{J_{\boldsymbol{A}}\} &= n\Psi^{*} \frac{Q_{\boldsymbol{A}}}{n}  - \left(\frac{Q_{\boldsymbol{\check{A}}}}{n}\right)\frac{n}{2}(\boldsymbol{\omega}_{(d),\boldsymbol{A}} - \boldsymbol{w}_{(d)}^{*})^{T} H_{\Psi}^{*}  (\boldsymbol{\omega}_{(d),\boldsymbol{A}} - \boldsymbol{w}_{(d)}^{*}).
\end{align}

The first lemma can be considered a corollary to \cite{Lane:Adap:2019} Theorem 1.
\begin{lemma}\label{lem:Info}
Under the present conditions $\omega_{\boldsymbol{a}_{i}^{*}} - w_{i}^{*} = O_{p}(n^{-1/2})$ and $\omega_{\check{\boldsymbol{a}}_{i}} - w_{i}^{*} = O_{p}(n^{-1})$ for $i=1,\ldots,d$.
\end{lemma}
For $\omega_{\boldsymbol{a}_{i}^{*}}$ the conditions are more restrictive than the conditions in \cite{Lane:Adap:2019} Theorem 1 and thus $\omega_{\boldsymbol{a}_{i}^{*}} - w_{i}^{*} = O_{p}(n^{-1/2})$. For $\omega_{\check{\boldsymbol{a}}_{i}}$ all conditions are satisfied by Lemma 2 in \cite{Lane:Adap:2019} trivially except the condition that $\check{x}(j)$ occurs at a point where $w_{ij}'>0$, where $w_{ij}'$ is defined as
\begin{align}
w_{ij}' &= w_{i}^{*} + Q_{\boldsymbol{A}(j-1)}\{w_{i}^{*}-\omega_{\boldsymbol{a}_{i}(j-1)}\} = w_{i}^{*}(1 + Q_{\boldsymbol{A}(j-1)}) - Q_{\boldsymbol{A}(j-1)}\omega_{\boldsymbol{a}_{i}(j-1)} \\
&> Q_{\boldsymbol{A}(j-1)}(w_{i}^{*} - \omega_{\boldsymbol{a}_{i}(j-1)}).
\end{align}
In other words, if $w_{i}^{*} > \omega_{\boldsymbol{a}_{i}(j-1)}$ then $w_{ij}'>0$. By definition, the ROAD only searches for the $j$th optimal design point over the set of $x^{*}$ such that $w_{i}^{*} > \omega_{\boldsymbol{a}_{i}(j-1)}$. Therefore, the conditions of \cite{Lane:Adap:2019} Theorem 1 are satisfied for the ROAD and $\omega_{\check{\boldsymbol{a}}_{i}} - w_{i}^{*} = O_{p}(n^{-1})$. $\square$

This below lemma address the first term in \ref{eq:EffOAD}. 
\begin{lemma} \label{lem:EQ} Under the present conditions
${\rm{E}}[Q_{\mathcal{A}}/n] = h + o(n^{-1})$, where $\mathcal{A}$ can be either $\mathcal{\check{A}}$ or $\mathcal{A^{*}}$
\end{lemma}
Proof. In the ROAD the ancillary statistics are not independent and thus a significant amount of care is needed to show certain results that can be easily shown for a fixed design.  Following a ROAD $n_{i}$ is not fixed as it is for the FOD. Instead, in a ROAD the observed $n_{i}$ is a random variable and is denoted $N_{i}(n)$. Further, note that $\boldsymbol{\check{a}_{i}}$ is not a sequence of independent variables, once again this differs from fixed designs. The notation of the derivatives of the log likelihood, in the proof of Theorem \ref{thm:MSE} are altered to reflect this as $l_{N_{i}(n)}^{(\cdot k)}(\boldsymbol{\varepsilon_{i}})$. 

The following lemma gives the asymptotic behavior of the higher order derivatives of log likelihood functions evaluated at the MLE.
\begin{lemma} \label{lem:LikLim} Under the present conditions $n^{-1} l_{\boldsymbol{\check{a}_{i}}}^{(\cdot k)} \rightarrow w_{i}^{*}\mu_{k}$ in probability as $n\rightarrow \infty$ for $k=1,\ldots,4$, where $\mu_{k} = {\rm{E}}[l_{0}^{(\cdot k)}(\varepsilon_{ij})]$.
\end{lemma}
The proof of this lemma is given in the supplemental materials.

Returning to the proof of Lemma \ref{lem:EQ}. After re-arranging a three term Taylor expansion for $\boldsymbol{i}_{\boldsymbol{\check{a}_{i}}}$ we get
\begin{align} \label{eq:InfoApprox}
\boldsymbol{i}_{\boldsymbol{\check{a}_{i}}} =  -l_{N_{i}(n)}^{(\cdot 2)}(\boldsymbol{\varepsilon_{i}}) - e_{i}l_{\boldsymbol{\check{a}_{i}}}^{(\cdot 3)} +e_{i}^{2}\frac{1}{2}l_{\boldsymbol{\check{a}_{i}}}^{(\cdot 4)} + o_{p}(1).
\end{align}
Now from Lemmas \ref{lem:efro:hink} and \ref{lem:LikLim} we can write the first two conditional moments as
\begin{align}
n{\rm{E}}[e_{i}|\boldsymbol{\check{a}_{i}}] = -\frac{\mu_{3}}{2w_{i}\mu^{2}} + o_{p}(1) \quad \mbox{and} \quad
n{\rm{E}}[e_{i}^{2}|\boldsymbol{\check{a}_{i}}] &= (2w_{i}\mu)^{-1}  + o_{p}(1).
\end{align}
The above holds for all $i=1,\ldots,d$. Taking the conditional expectation of both sides of \eqref{eq:InfoApprox} and recognizing that $Q_{\boldsymbol{\check{A}}} = \sum_{i=1}^{d} \mu^{-1}\boldsymbol{i}_{\boldsymbol{\check{a}_{i}}}$
\begin{align}
n^{-1}\sum_{i=1}^{d}{\rm{E}}[Q_{\mathcal{\check{A}}}] &= -(n\mu)^{-1}{\rm{E}}[l_{n}^{(\cdot 2)}(\boldsymbol{\varepsilon})] + \frac{d}{\mu}\left(\frac{\mu_{3}^{2}}{2\mu^{2}} + \frac{\mu_{4}}{2\mu}\right) + o(n^{-1}) = h + o(n^{-1}).
\end{align}
The same steps with $N_{i}(n)$ fixed can be used to show the lemma holds for fixed designs. $\square$ 

Return to equation \eqref{eq:EffOAD} for $J_{\boldsymbol{\check{A}}}$. Lemma \ref{lem:Info} ensures that the second and third terms of the right hand side are $O_{p}(n^{-1})$ and $o_{p}(n^{-1})$, respectively. Therefore $E[\Psi\{J_{\mathcal{\check{A}}}\}] = h\Psi^{*}n + o(1)$ as stated.

Consider equation \eqref{eq:EffOAD} for $\Psi\{J_{\boldsymbol{A}^{*}}\}$. Lemma \ref{lem:EQ} ensures that ${\rm{E}}[Q_{\mathcal{A}^{*}}/n] = h + o(n^{-1})$. The next lemma describes the asymptotic distribution of  $\boldsymbol{\omega}_{(d),\boldsymbol{A^{*}}}$. Note, this result holds only for fixed designs.
\begin{lemma} \label{lem:omegaDist} Under the present conditions
$\sqrt{n} (\boldsymbol{\omega}_{(d),\mathcal{A^{*}}} - \boldsymbol{w}_{(d)}^{*}) = N[0,\gamma^{2}V_{\Psi}^{*} ]$.
\end{lemma}
Proof. Under the assumed conditions
$\sqrt{n}(q_{\boldsymbol{a}_{i}}/n - w_{i}^{*}) \rightarrow N(0,w_{i}^{*}\gamma^{2})$ for $i=1,\ldots,d$ in distribution as $n\rightarrow \infty$ see \cite{Efro:Hink:Asse:1978}. The independence assumption implies that $n^{-1/2}(\boldsymbol{q_{A^{*}}}/n - \boldsymbol{w}^{*})$ is a vector of independent random variables with an approximate normal distribution with mean $\boldsymbol{0}$ and variance $\gamma^{2}W^{*}$, where $\boldsymbol{q}_{\boldsymbol{A}^{*}} =  (\boldsymbol{q}_{\boldsymbol{a}_{1}^{*}},\ldots,\boldsymbol{q}_{\boldsymbol{a}_{d}^{*}})^{T}$.

Let $\boldsymbol{g}(\boldsymbol{q_{A^{*}}}) = \boldsymbol{q}_{(d)A^{*}}(\boldsymbol{1}_{d}^{T}\boldsymbol{q_{A^{*}}})^{-1}$, where $\boldsymbol{q}_{(d)A^{*}}$ indicates that the $d$th element has been removed and let $\dot{\boldsymbol{g}}(\boldsymbol{q_{A^{*}}})$ represent the derivative with respect to $\boldsymbol{q_{A^{*}}}$ and note that
\begin{align}
\sqrt{n} (\boldsymbol{\omega}_{(d),\boldsymbol{A^{*}}} - \boldsymbol{w}_{(d)}^{*}) &= \sqrt{n}[\boldsymbol{q}_{(d),A^{*}}(\boldsymbol{1}_{d}^{T}\boldsymbol{q_{A^{*}}})^{-1} - \boldsymbol{w}_{(d)}^{*}] = \sqrt{n}[\boldsymbol{g}(\boldsymbol{q_{A^{*}}}) - \boldsymbol{w}_{(d)}^{*}]^{T},
\end{align}
Note $\boldsymbol{g}(\boldsymbol{w}^{*}) = \boldsymbol{w}_{(d)}^{*}$ and $\dot{\boldsymbol{g}}(\boldsymbol{w}^{*}) = (I_{d-1} - J_{\Psi}^{*},w_{d}^{*}\boldsymbol{1}_{d-1}^{T})$, where $J_{\Psi}^{*} = J_{d-1}diag(\boldsymbol{w}_{(d)}^{*})$, $I_{d-1}$ is the identity matrix and $J_{d-1}$ is a matrix where each element is equal to 1. The delta method implies the result of the lemma after some basic algebra. $\square$

The second term in the right hand side of \eqref{eq:EffOAD} for $\Psi\{J_{\boldsymbol{A^{*}}}\}$  is a quadratic form of a random variable $\boldsymbol{\omega}_{(d),\boldsymbol{A^{*}}}$. We consider only concave functions $\Psi$, as a consequence the negative of the Hessian matrix, $H_{\Psi}^{*}$ is positive definite. Thus Lemma \ref{lem:omegaDist} ensures
\begin{align}
n{\rm{E}}\left[(\boldsymbol{\omega}_{(d),\mathcal{A^{*}}} - \boldsymbol{w}_{(d)}^{*})^{T} H_{\Psi}^{*}  (\boldsymbol{\omega}_{(d),\mathcal{A^{*}}} - \boldsymbol{w}_{(d)}^{*})/2\right] \rightarrow \gamma^{2}{\rm{tr}}(H_{\Psi}^{*}V_{\Psi}^{*})/2.
\end{align}
Therefore we can write $E[\Psi\{J_{\mathcal{A}^{*}}\}] = h\Psi^{*}\left[n - \gamma^{2}R_{\Psi}^{*}\right] + o(1)$ as stated. 

\subsection{Proof of Theorem \ref{thm:MSEopt}}
The conditions for this theorem are the same as those for Theorem \ref{thm:opt} with the additional condition that $E[e_{i}^{4}]<\infty$. For the proof of this theorem it is required to calculate the limit of the $O_{p}(n^{-1})$ term of the conditional MSE which is given in the following Lemma
\begin{lemma} Under the present conditions
\begin{align}
    {\rm{MSE}}[\hat{\boldsymbol{\beta}}|\mathcal{A}] = J_{\mathcal{A}}^{-1} + n^{-2}G'(\boldsymbol{w},\boldsymbol{\mu}) + o_{p}(n^{-2}),
\end{align}
where $G'$ is a finite matrix and $\boldsymbol{\mu} = (\mu,\mu_{3},\mu_{4})^{T}$. 
\end{lemma}
The proof of this lemma along with details regarding $G'$ are given in the supplemental materials.

Letting $c_{n} = n^{-1}$ then a Taylor expansion around $c_{n}=0$ yields
\begin{align}
n^{-1}\Psi({\rm{MSE}}[\hat{\boldsymbol{\beta}}|\mathcal{A}]^{-1}) &= \Psi\{[nJ_{\mathcal{A}}^{-1} + c_{n}G'(\boldsymbol{w},\boldsymbol{\mu}) + o_{p}(n^{-1}) ]^{-1}\} \\
&= n^{-1}\Psi(J_{\mathcal{A}}) + c_{n}U_{\mathcal{A}} + o_{p}(n^{-1}),
\end{align}
where $U_{\boldsymbol{A}} = (\partial/\partial c_{n})\Psi\{[nJ_{\boldsymbol{A}}^{-1} + c_{n}G'(\boldsymbol{w},\boldsymbol{\mu}) ]^{-1}\}_{c_{n}=0}$. The above holds for any arbitrary $\mathcal{A}$. Specifically, it holds for $\mathcal{A}$ equal to $\mathcal{\check{A}}$ or $\mathcal{A}^{*}$. Therefore we can write
\begin{align} \label{eq:MSEdiff}
    \Psi({\rm{MSE}}[\hat{\boldsymbol{\beta}}|\mathscr{\check{A}}]^{-1})  - \Psi({\rm{MSE}}[\hat{\boldsymbol{\beta}}|\mathcal{A^{*}}]^{-1}) &= \Psi(J_{\mathcal{\check{A}}}) - \Psi(J_{\mathcal{A^{*}}}) + c_{n} \left[U_{\mathcal{\check{A}}} - U_{\mathcal{A^{*}}} \right] \\
    &\quad\quad+ o_{p}(1).
\end{align}
From Theorem \ref{thm:opt} it is known that $J_{\mathcal{\check{A}}} - J_{\mathcal{A^{*}}} = O_{p}(n^{-1/2})$; therefore,
\begin{align}
E[ \Psi({\rm{MSE}}[\hat{\boldsymbol{\beta}}|\mathscr{\check{A}}]^{-1})  - \Psi({\rm{MSE}}[\hat{\boldsymbol{\beta}}|\mathcal{A^{*}}]^{-1})] = \gamma^{2}tr(H_{\Psi}^{*}V_{\Psi}^{*})/2 + o(1).
\end{align}
The term $tr(H_{\Psi}^{*}V_{\Psi}^{*})$ is shown to be greater than or equal to 0 if $\gamma^{2}>0$ and $d>1$ in the proof of Corollary \ref{cor:opt} which concludes the proof.
%%%%%%%%%%%%%%%%%%%%%%%%%%%%%%%%%%%%%%%%%%%%%%
%% Please use \tableofcontents for articles %%
%% with 50 pages and more                   %%
%%%%%%%%%%%%%%%%%%%%%%%%%%%%%%%%%%%%%%%%%%%%%%
%\tableofcontents

%%%%%%%%%%%%%%%%%%%%%%%%%%%%%%%%%%%%%%%%%%%%%%
%%%% Main text entry area:

%%%%%%%%%%%%%%%%%%%%%%%%%%%%%%%%%%%%%%%%%%%%%%
%% Single Appendix:                         %%
%%%%%%%%%%%%%%%%%%%%%%%%%%%%%%%%%%%%%%%%%%%%%%
%\begin{appendix}
%\section*{???}%% if no title is needed, leave empty \section*{}.
%\end{appendix}
%%%%%%%%%%%%%%%%%%%%%%%%%%%%%%%%%%%%%%%%%%%%%%
%% Multiple Appendixes:                     %%
%%%%%%%%%%%%%%%%%%%%%%%%%%%%%%%%%%%%%%%%%%%%%%
%\begin{appendix}
%\section{???}
%
%\section{???}
%
%\end{appendix}

%%%%%%%%%%%%%%%%%%%%%%%%%%%%%%%%%%%%%%%%%%%%%%
%% Support information (funding), if any,   %%
%% should be provided in the                %%
%% Acknowledgements section.                %%
%%%%%%%%%%%%%%%%%%%%%%%%%%%%%%%%%%%%%%%%%%%%%%
% \section*{Acknowledgements}
% The authors would like to thank ...
% 
% The first author was supported by ...
% 
% The second author was supported in part by ...
 
%%%%%%%%%%%%%%%%%%%%%%%%%%%%%%%%%%%%%%%%%%%%%%
%% Supplementary Material, if any, should   %%
%% be provided in {supplement} environment  %%
%% with title inside \textbf{} and short    %%
%% description below.                       %%
%%%%%%%%%%%%%%%%%%%%%%%%%%%%%%%%%%%%%%%%%%%%%%
\begin{supplement}
\begin{center}
%The title should be centred and in bold letters. It should be informative but not too long (preferably no more than two lines).
\textbf{Supplemental Materials for Efficiency of Observed Information Adaptive Designs}
\end{center}

\section{Gamma Hyperbola Linear Regression}

In this section we show the gamma hyperbola model can be written as a linear regression model. The model is derived for a fixed design. We begin by stating the distribution of $(s,t) = (z_{1}e^{\eta},z_{2}e^{-\eta})$, which is the product of two independent gamma random variables, i.e.
\begin{align}
f_{\eta}(s,t) = e^{-(se^{\eta} + te^{-\eta})} \frac{1}{\Gamma[v]^{2}}(st)^{v-1}
\end{align}
where $\eta = \beta^{T}f_{x}(x)$. Let $a = \sqrt{st}$ and recall $\frac{1}{2}\log[s/t]$ then using a change of variable we get
\begin{align}
f_{\eta}(y,a) &=  \frac{1}{\Gamma[\beta]^{2}}a^{v-1} e^{-a (  e^{-(y - \eta)} + e^{y - \eta} ) } \\
&= \frac{1}{\Gamma[\beta]^{2}}a^{v-1} e^{2a \mbox{Cosh}(y - \eta) }.
\end{align}
Using the above expression we can now see that the location family condition that $f_{0}(\varepsilon) = f_{0}(y-\eta) = f_{\eta}(y)$ is satisfied. 

\section{Non-Optimal Fixed Designs}
Let $\xi=\{(x_{i},w_{i})\}_{i=1}^{d}$ be any arbitrary design in $\Xi_{\Delta}$. Here we are interested in $\xi$ such that $\Psi[M(\xi)] = \Psi_{\Delta}^{*} + c_{\xi}n^{-\delta}$, where $c_{\xi} > 0$ is a finite constant $\delta\in[0,1)$ for any design.  The term $\delta$ defines the degree to which $\xi$ is approximately a FOD. Theorem \ref{thm:MSEopt} has already addressed $\delta\ge 1$.

For $\delta<1$, a Taylor expansion of $\boldsymbol{i}_{\boldsymbol{A}}$ around $n\mu\boldsymbol{w^{*}}$, where $\boldsymbol{i}_{\boldsymbol{A}} = (\boldsymbol{i}_{\boldsymbol{a_{1}}},\ldots,\boldsymbol{i}_{\boldsymbol{a_{d}}})^{T}$.
\begin{align}
    \Psi(J_{\boldsymbol{A}}) &= \Psi[M(\xi)] +  (\boldsymbol{i}_{\boldsymbol{A}} - n\mu\boldsymbol{w^{*}})^{T}\nabla\Psi[M(\xi)] \\
    &\quad\quad- \frac{1}{2}(\boldsymbol{i}_{\boldsymbol{A}} - n\mu\boldsymbol{w^{*}})^{T}H_{\Psi}(\boldsymbol{i}_{\boldsymbol{A}} - n\mu\boldsymbol{w^{*}}) + o_{p}(1).
\end{align}
The same steps to show \eqref{eq:ExpNi} for random sample sizes still hold for fixed sample sizes; therefore, 
\begin{align} \label{eq:JAineq}
   E[\Psi(J_{\mathcal{A}})] = n\Psi[M(\xi)] + O_{p}(1).
\end{align}
This implies that
\begin{align} 
    \{E[\Psi(J_{\mathcal{A}^{*}})] - E[\Psi(J_{\mathcal{A}})]\} = c_{\xi}n^{1-\delta} + O_{p}(1),
\end{align}
where $J_{\boldsymbol{A}^{*}}$ be the observed Fisher information with design satisfying $\Psi[M(\xi_{\Psi}^{*})]= \Psi_{\Delta}^{*}[1+O(n^{-1})]$ and $J_{\boldsymbol{A}}$ is the observed Fisher information from any design with $\delta<1$. The first term in the right hand side goes to infinity as $n\rightarrow$. Equation \eqref{eq:MSEdiff} can be used to show that this is also true for the conditional MSE as claimed in the main text. Therefore, no design with $\delta<1$ is asymptotically more efficient that the FOD. 

\section{Gradient of $\Psi$}
To begin assume that $\xi_{\Psi}^{*}$ is a continuous optimal design. The gradient of $\Psi$ must be found under the constraint $\omega_{\boldsymbol{a}_{d}} = 1 - \sum_{i=1}^{d-1}\omega_{\boldsymbol{a}_{i}}$. Under this constraint the total derivative treating $\omega_{\boldsymbol{a}_{k}}, k\ne i = 1,\ldots,d-1$ as fixed is
\begin{align}
\frac{d}{d \omega_{\boldsymbol{a}_{i}} } \Psi\{M(\tau_{\boldsymbol{A}})\} &= \frac{\partial}{\partial \omega_{\boldsymbol{a}_{i}} } \Psi\{M(\tau_{\boldsymbol{A}})\} + \frac{\partial}{\partial \omega_{\boldsymbol{a}_{d}} } \Psi\{M(\tau_{\boldsymbol{A}})\}  \frac{d \omega_{\boldsymbol{a}_{d}} }{d \omega_{\boldsymbol{a}_{i}} } \\
&= -f^{T}(x_{i}^{*})\frac{\partial \Psi}{\partial M}f^{T}(x_{i}^{*}) -f^{T}(x_{d}^{*})\frac{\partial \Psi}{\partial M}f^{T}(x_{d}^{*}) \frac{d \omega_{\boldsymbol{a}_{d}} }{d \omega_{\boldsymbol{a}_{i}} }
\end{align}
for $i=1,\ldots,d-1$. From the general equivalence theorem $f^{T}(x_{i}^{*})\frac{\partial \Psi}{\partial M}f^{T}(x_{i}^{*})|_{M = M_{\Psi}^{*}} = {\rm{tr}}(M\frac{\partial \Psi}{\partial M})|_{M = M_{\Psi}^{*}}$ for all optimal design points. Further, taking the derivative of the constraint yields
$d \omega_{\boldsymbol{a}_{d}} = -d \omega_{\boldsymbol{a}_{i}}$.Therefore,
\begin{align} \label{eq:dPsi0}
\left[\frac{d}{d \omega_{\boldsymbol{a}_{i}} } \Psi\{M(\tau_{\boldsymbol{A}})\}\right]_{M(\tau_{\boldsymbol{A}}) = M_{\Psi}^{*}} &= 0
\end{align}
which implies the gradient is a vector of zeros under the constraint. For $\xi_{\Psi}^{*}$ such that $\Psi[M(\xi_{\Psi}^{*})]= \Psi_{\Delta}^{*}[1+O(n^{-1})]$ it is straightforward to show that $\nabla  \Psi\{M(\tau_{\boldsymbol{A}})\} = O_{p}(n^{-1})$.

\section{Additional Technical Details} \label{suppsec:Tech}

In this section the supporting arguments for the additional theoretical results are presented.

\subsection{Proof of Corollary \ref{cor:efro:hink}}
First, consider the sufficiency result. It is well known that the likelihood following an adaptive design does not change the form of the likelihood \cite{Ford:Titt:Wu:infe:1985}. Therefore, $L(\boldsymbol{\beta}|\boldsymbol{y}) \propto \prod_{i=1}^{d} g_{\boldsymbol{a_{i}}}[\hat{\eta}_{i} - \boldsymbol{\beta}^{T}f_{x}(x_{i})]$ is still the likelihood following the ROAD. However, note that $\boldsymbol{\hat{\eta}}$ is a function of $\boldsymbol{N(n)} = [N_{1}(n),\ldots,N_{d}(n)]$. Observing the likelihood and we can see that $(\boldsymbol{\hat{\eta}},\boldsymbol{N(n)},\mathcal{\check{A}})$ is a sufficient statistic. Now from the proof of Theorem \ref{thm:ancillary} $\boldsymbol{N(n)}$ and $\mathcal{\check{A}}$ are $\mathscr{\check{A}}$ measurable. Therefore, $(\boldsymbol{\hat{\eta}},\mathscr{\check{A}})$ is a sufficient statistic. Finally, since $\boldsymbol{\hat{\beta}}$ given $\mathscr{\check{A}}$ is an injection of $\boldsymbol{\hat{\eta}}$ it follows that $(\boldsymbol{\hat{\beta}},\mathscr{\check{A}})$ is a sufficient statistic. 

The steps used to prove 
\begin{align}
n{\rm{MSE}}[\hat{\boldsymbol{\beta}}|\mathcal{A}] &=  nJ_{\mathcal{A}}^{-1}\left[1 +  O_{p}(n^{-1})\right].
\end{align}
in Theorem \ref{thm:MSE} were all based on the conditional distribution of the MLE. Therefore, by Theorem \ref{thm:ancillary} the exact same steps are still valid following the ROAD with $\mathcal{A}$ replaced with $\mathscr{\check{A}}$. This directly implies that
\begin{align}
n{\rm{MSE}}[\hat{\boldsymbol{\beta}}|\mathscr{\check{A}}] &=  nJ_{\mathcal{\check{A}}}^{-1}\left[1 +  O_{p}(n^{-1})\right].
\end{align}
All that remains to be shown is that $n^{-1}[J_{\mathcal{A}} - \mathcal{F}(\check{\xi}_{\Psi})] = O_{p}(n^{-1/2})$ following the ROAD procedure. Write
\begin{align}
    J_{\mathcal{A}} - \mathcal{F}(\check{\xi}_{\Psi}) = \sum_{i=1}^{d}[\boldsymbol{i}_{\boldsymbol{\check{a}_{i}}} - \mu N_{i}(n)]f_{x}(x_{i}^{*})f_{x}^{T}(x_{i}^{*}).
\end{align}
The matrices $f_{x}(x_{i}^{*})f_{x}^{T}(x_{i}^{*})$ are finite for all $i$; therefore, the above has the same order as
\begin{align}
    \sum_{i=1}^{d}[\boldsymbol{i}_{\boldsymbol{\check{a}_{i}}} - \mu N_{i}(n)] = \mu(Q_{\boldsymbol{A}} -  n).
\end{align}
Recalling that $Q_{\boldsymbol{\check{A}}} = \sum_{i=1}^{d} \mu^{-1}\boldsymbol{i}_{\boldsymbol{\check{a}_{i}}}$ and using the expansion in \eqref{eq:InfoApprox} we can write 
\begin{align}
\mu\sqrt{n}(Q_{\boldsymbol{A}}/n -  1) = \sqrt{n}(-\ddot{l}_{n}(\boldsymbol{\varepsilon})/n - 1) + O_{p}(n^{-1/2}).
\end{align}
The central limit theorem ensures that the right hand side of the above has a limiting normal distribution; therefore $(Q_{\boldsymbol{A}}/n -  1) = O_{p}(n^{-1/2})$ which implies the stated result.

\subsection{Proof of Lemma \ref{lem:LikLim} }

The following lemma describes the limiting behavior of the random sample sizes of the ROAD on each optimal support point.
\begin{lemma}\label{lem:Ni} Under the conditions of Theorem \ref{thm:opt}
\begin{align}
N_{i}(n)/n \rightarrow w_{i}^{*}
\end{align}
in probability as $n\rightarrow\infty$ for $i=1,\ldots,d$.
\end{lemma}
Proof. From \eqref{eq:InfoApprox}
\begin{align} 
n^{-1}{\rm{E}}[\boldsymbol{i}_{\boldsymbol{\check{a}_{i}}}] &= n^{-1}E\left[-l_{N_{i}(n)}^{(\cdot 2)}(\boldsymbol{\varepsilon_{i}}) - e_{i}l_{\boldsymbol{\check{a}_{i}}}^{(\cdot 3)} +e_{i}^{2}\frac{1}{2}l_{\boldsymbol{\check{a}_{i}}}^{(\cdot 4)}\right] +  o(n^{-1}) \\
&= -n^{-1}{\rm{E}}[{\rm{E}}[l_{N_{i}(n)}^{(\cdot 2)}(\boldsymbol{\varepsilon_{i}})|N_{i}(n)] - n^{-1}E\left[ l_{\boldsymbol{\check{a}_{i}}}^{(\cdot 3)}E[e_{i}|\boldsymbol{\check{a}_{i}}]\right] \\
&\quad\quad+E\left[\frac{1}{2}l_{\boldsymbol{\check{a}_{i}}}^{(\cdot 4)}E[e_{i}^{2}|\boldsymbol{\check{a}_{i}}]\right]+  o(n^{-1}) 
\end{align}
From Lemma \ref{lem:efro:hink} the first two conditional moments of $e_{i}$ are $O_{p}(n^{-1})$; therefore,
\begin{align} \label{eq:ExpNi}
{\rm{E}}[\boldsymbol{i}_{\boldsymbol{\check{a}_{i}}}/n] - n^{-1}{\rm{E}}[{\rm{E}}[l_{N_{i}(n)}^{(\cdot 2)}(\boldsymbol{\varepsilon_{i}})|N_{i}(n)]={\rm{E}}[\boldsymbol{i}_{\boldsymbol{\check{a}_{i}}}/n] - \mu {\rm{E}}[N_{i}(n)/n] = o(n^{-1/2}).
\end{align}
Next consider the variance
\begin{align}
n^{-2}{\rm{Var}}[\boldsymbol{i}_{\boldsymbol{\check{a}_{i}}}] &=  n^{-1}{\rm{Var}}[l_{N_{i}(n)}^{(\cdot 2)}(\boldsymbol{\varepsilon_{i}})] + o(1) \\
&= n^{-2}{\rm{E}}[{\rm{Var}}[l_{N_{i}(n)}^{(\cdot 2)}(\boldsymbol{\varepsilon_{i}})|N_{i}(n)]] + n^{-2}{\rm{Var}}[{\rm{E}}[l_{N_{i}(n)}^{(\cdot 2)}(\boldsymbol{\varepsilon_{i}})|N_{i}(n)]] + o(1) \\
&= {\rm{Var}}[N_{i}(n)/n] + o(1) \label{eq:VarNi}
\end{align}
since ${\rm{Var}}[l_{N_{i}(n)}^{(\cdot 2)}(\boldsymbol{\varepsilon_{i}})|N_{i}(n)]<\infty$ by assumption. Chebychev's inequality ensures $N_{i}(n)/n$ will converge in probability to its expectation if ${\rm{Var}}[N_{i}(n)/n]\rightarrow 0$ as $n\rightarrow \infty$. From \eqref{eq:VarNi} this is equivalent to $n^{-2}{\rm{Var}}[\boldsymbol{i}_{\boldsymbol{a_{i}}}]\rightarrow 0$.

It was shown in the proof of \ref{cor:efro:hink} that  $Q_{\boldsymbol{A}}/n = (n\mu)^{-1}\sum_{i}\boldsymbol{i}_{\boldsymbol{\check{a}_{i}}}$ is has a limiting normal distribution. This implies that $\boldsymbol{i}_{\boldsymbol{\check{a}_{i}}}/n$ has a finite variance fr all $i$ and thus by Chebychev's inequality it also converges in probability to its mean. Recall that $\omega_{\boldsymbol{a}_{i}^{*}} = \boldsymbol{i}_{\boldsymbol{\check{a}_{i}}}/\sum_{i}\boldsymbol{i}_{\boldsymbol{\check{a}_{i}}}$ then Lemma \ref{lem:Info} implies that
\begin{align}
{\rm{E}}[\boldsymbol{i}_{\boldsymbol{\check{a}_{i}}}/n] \rightarrow \mu w_{i}^{*}
\end{align}
in probability as $n\rightarrow \infty$. Therefore,
\begin{align}
N_{i}(n)/n \rightarrow w_{i}^{*}
\end{align}
in probability as $n\rightarrow \infty$ as stated. $\square$

Using steps similar to those used to show \eqref{eq:ExpNi} and \eqref{eq:VarNi} it can be shown that
\begin{align}
n^{-1}{\rm{E}}[l_{\boldsymbol{\check{a}_{i}}}^{(\cdot k)}] &= n^{-1}\mu_{k} {\rm{E}}[N_{i}(n)] + o(1) \\
n^{-2}{\rm{{\rm{Var}}}}[l_{\boldsymbol{\check{a}_{i}}}^{(\cdot k)}] &= {\rm{Var}}[N_{i}(n)/n] + o(1)
\end{align}
for $i=3,4$. The lemma follows from Lemma \ref{lem:Ni} and Chebychev's inequality. $\square$

\subsection{Proof of Corollary \ref{cor:opt}}
Theorem \ref{thm:opt} directly implies
\begin{align} \
\Psi\mbox{-Eff}_{\rm{CI}} = \left[1 - \frac{\gamma^{2}}{n}R_{\Psi}^{*}\right]^{-1} + o(1).
\end{align}
To show that $\Psi\mbox{-Eff}_{\rm{CI}}\ge1$ note that \cite{Fang:Lopa:Feng:1994} show that lower bound on the trace can be written,
\begin{align}
R_{\Psi}^{*} = \frac{\gamma^{2}}{2}{\rm{tr}}(H_{\Psi}^{*}V_{\Psi}^{*}) \ge \frac{\gamma^{2}}{2}\lambda_{n}(H_{\Psi}^{*}){\rm{tr}}(V_{\Psi}^{*}) 
\end{align}
where $\lambda_{n}(H_{\Psi}^{*})>0$ is the smallest eigenvalue of $H_{\Psi}^{*}$ which is positive definite. Further, after some basic algebra it can be shown that
\begin{align}
    {\rm{tr}}(V_{\Psi}^{*}) = (d-1)\sum_{i=1}^{d-1}w_{i}^{3} + \sum_{i=1}^{d-1}w_{i}(1-w_{i}) > 0
\end{align}
if $d>1$. Therefore, $R_{\Psi}^{*}>0$ if $d>1$.

%For the only if direction, note that $\boldsymbol{\omega}_{\boldsymbol{A}} = 1$, is a constant. %Therefore, the expansion in \eqref{eq:ObsExp} reduces to
%\begin{align}
%\Psi\{M(\tau_{\boldsymbol{A}})\} &= \Psi\{M(\xi^{*})\}.
%\end{align}
%This above is true for both the ROAD and FOD and thus $R_{\Psi}^{*}$ is 0 if $d=1$. This concludes the %proof of the if and only if statement.

\subsection{Proof of Lemma \ref{thm:MSEopt}}
The following lemma considers the fourth conditional moment of $e_{i}$.
\begin{lemma} \label{lem:e4} Under the present conditions the following hold
\begin{align}
E[e_{i}^{4}|\boldsymbol{a_{i}}] &= 3 \{\boldsymbol{i}_{\boldsymbol{a_{i}}}\}^{-2} + o_{p}(n^{-2})
\end{align}
for $i=1,\ldots,d$.
\end{lemma}
The proof for is obtained using very similar steps as those used in \cite{Efro:Hink:Asse:1978} and is omitted.

Next, expanding the conditional MSE yields
\begin{align}
        (\boldsymbol{\hat{\beta}} - \boldsymbol{\beta})(\boldsymbol{\hat{\beta}} - \boldsymbol{\beta})^{T} &= J_{\boldsymbol{A}}^{-1}F^{T} \left[I_{\boldsymbol{A}}\boldsymbol{e}\boldsymbol{e}^{T} I_{\boldsymbol{A}} -  I_{\boldsymbol{A}}\boldsymbol{e}\boldsymbol{\hat{e}}^{T}diag(\boldsymbol{\hat{e}}) K_{\boldsymbol{A}} \right. \\
        &\quad\quad\left.+\frac{1}{4} K_{\boldsymbol{A}}diag(\boldsymbol{\hat{e}})\boldsymbol{\hat{e}}\boldsymbol{\hat{e}}^{T}diag(\boldsymbol{\hat{e}}) K_{\boldsymbol{A}}\right.\\
        &\quad\quad\left.-\frac{1}{3} I_{\boldsymbol{A}}\boldsymbol{e}\boldsymbol{\hat{e}}^{T}diag(\boldsymbol{\hat{e}}^{T}\boldsymbol{\hat{e}})L_{\boldsymbol{A}}  \right]FJ_{\boldsymbol{A}}^{-1} + o_{p}(n^{-2}).
\end{align}
Now using the equation \eqref{eq:resid} we can write
\begin{align}
    \boldsymbol{\hat{e}}^{T}diag(\boldsymbol{\hat{e}}) &= \left[(I_{d} - P_{\boldsymbol{A}})\boldsymbol{e} -\frac{1}{2} FJ_{\boldsymbol{A}}^{-1} F^{T} K_{\boldsymbol{A}} diag(\boldsymbol{\hat{e}})\boldsymbol{\hat{e}}\right]^{T}diag\{(I_{d} - P_{\boldsymbol{A}})\boldsymbol{e} \\
    &\quad\quad-\frac{1}{2} FJ_{\boldsymbol{A}}^{-1} F^{T} K_{\boldsymbol{A}} diag(\boldsymbol{\hat{e}})\boldsymbol{\hat{e}}\} \\
    &= \boldsymbol{e}^{T}(I_{d} - P_{\boldsymbol{A}})^{T}diag\{\boldsymbol{e}(I_{d} - P_{\boldsymbol{A}})\} \\
    &\quad\quad -\frac{1}{2} \boldsymbol{e}^{T}(I_{d} - P_{\boldsymbol{A}})^{T}diag\{FJ_{\boldsymbol{A}}^{-1} F^{T} K_{\boldsymbol{A}} diag(\boldsymbol{\hat{e}})\boldsymbol{\hat{e}}\}\\
    &\quad\quad -\frac{1}{2} \boldsymbol{\hat{e}}^{T}diag(\boldsymbol{\hat{e}})K_{\boldsymbol{A}}FJ_{\boldsymbol{A}}^{-1}F^{T}diag\{(I_{d} - P_{\boldsymbol{A}})\boldsymbol{e}\} +O_{p}(n^{-2})
\end{align}
Now using the above and recalling that $\boldsymbol{\hat{e}} = (I_{d} - P_{\boldsymbol{A}})\boldsymbol{e} + O_{p}(n^{-1})$ we get
\begin{align}
       (\boldsymbol{\hat{\beta}} - \boldsymbol{\beta})(\boldsymbol{\hat{\beta}} - \boldsymbol{\beta})^{T} &= J_{\boldsymbol{A}}^{-1}F^{T} \left[I_{\boldsymbol{A}}\boldsymbol{e}\boldsymbol{e}^{T} I_{\boldsymbol{A}} - I_{\boldsymbol{A}}\boldsymbol{e}\boldsymbol{e}^{T}T_{\boldsymbol{A}} K_{\boldsymbol{A}} \right.\\
        &\quad\quad\left. + \frac{1}{2} I_{\boldsymbol{A}}\boldsymbol{e} \boldsymbol{e}^{T}(I_{d} - P_{\boldsymbol{A}})^{T} diag\{FJ_{\boldsymbol{A}}^{-1} F^{T} K_{\boldsymbol{A}} diag(\boldsymbol{e})\boldsymbol{e}\} K_{\boldsymbol{A}}
        \right.\\
        &\quad\quad\left. + \frac{1}{2} I_{\boldsymbol{A}}\boldsymbol{e} \boldsymbol{e}^{T}diag(\boldsymbol{e})K_{\boldsymbol{A}}FJ_{\boldsymbol{A}}^{-1}F^{T}diag\{(I_{d} - P_{\boldsymbol{A}})\boldsymbol{e}\} K_{\boldsymbol{A}}
        \right.\\
        &\quad\quad\left. + \frac{1}{4}
        K_{\boldsymbol{A}}diag(\boldsymbol{e})\boldsymbol{e}\boldsymbol{e}^{T}diag(\boldsymbol{e}) K_{\boldsymbol{A}} - \frac{1}{3} I_{\boldsymbol{A}}\boldsymbol{e}\boldsymbol{e}^{T}diag(\boldsymbol{e}^{T}\boldsymbol{e})L_{\boldsymbol{A}}  \right]FJ_{\boldsymbol{A}}^{-1} \\
        &\quad\quad + o_{p}(n^{-2}). \label{eq:MSEFullExp}
\end{align}
Now we consider the the bracketed terms in the above. The conditional expectation of the first term is
\begin{align}
    E[I_{\mathcal{A}}\boldsymbol{e}\boldsymbol{e}^{T}I_{\mathcal{A}}|\mathcal{A}] 
    &= I_{\mathcal{A}}\left\{I_{p} + \frac{1}{2}I_{\mathcal{A}}^{-1}\left[\frac{5}{4}I_{\mathcal{A}}^{-1}K_{\mathcal{A}}^{2} + L_{\mathcal{A}} \right] I_{\mathcal{A}}^{-1} \right\} + o_{p}(n^{-2}).
\end{align}
Next note that by Lemma \ref{lem:LikLim} the following hold
\begin{align}
    n^{-1}I_{\boldsymbol{A}} &= \mu W + o_{p}(1) \\
    n^{-1}K_{\boldsymbol{A}} &= \mu_{3}W + o_{p}(1)\\
    n^{-1}L_{\boldsymbol{A}} &= \mu_{4}W + o_{p}(1).
\end{align}
The above implies that $P_{\boldsymbol{A}} = P_{W} + o_{p}(1)$, where $P_{W} = FM^{-1}F^{T}W$. Using the above the second term can be written as
\begin{align}
    I_{\boldsymbol{A}}\boldsymbol{e}\boldsymbol{e}^{T}T_{\boldsymbol{A}} K_{\boldsymbol{A}} = n^{2}\mu\mu_{3}W\boldsymbol{e}\boldsymbol{e}^{T}(I_{d} - P_{\boldsymbol{W}})^{T}diag\{\boldsymbol{e}(I_{d} - P_{\boldsymbol{W}})\} W[1 + o_{p}(1)].
\end{align}
As shown in the proof of Theorem \ref{thm:MSE} the above is $O_{p}(n^{-2})$ and therefore the $o_{p}(1)$ can be ignored to obtain the required order of approximation. Using a similar method to the above the 2nd - 6th bracketed terms in \ref{eq:MSEFullExp} can be expressed as
\begin{align}
G(\boldsymbol{e}) &= n^{2}\left[-\mu\mu_{3}W\boldsymbol{e}\boldsymbol{e}^{T}(I_{d} - P_{\boldsymbol{W}})^{T}diag\{\boldsymbol{e}(I_{d} - P_{\boldsymbol{W}})\} W \right. \\
&\quad\quad\left. + \frac{1}{2}\mu_{3}^{2}W\boldsymbol{e} \boldsymbol{e}^{T}(I_{d} - P_{W})^{T} diag\{P_{W}diag(\boldsymbol{e})\boldsymbol{e}\} W \right.\\
&\quad\quad \left.+\frac{1}{2}\mu_{3}^{2}WP_{W}^{T}diag\{(I_{d} - P_{W})\boldsymbol{e}\}W \right.\\
&\quad\quad\left. +\frac{1}{4}\mu_{3}^{2}Wdiag(\boldsymbol{e})\boldsymbol{e}\boldsymbol{e}^{T}diag(\boldsymbol{e})W -\frac{1}{3}\mu\mu_{4}\boldsymbol{e}\boldsymbol{e}^{T}diag(\boldsymbol{e}^{T}\boldsymbol{e})\right].
\end{align}
respectively. The conditional expectation of $G(\boldsymbol{e})$ depends on the conditional moments of  $e_{i}^{r}e_{j}^{s}e_{k}^{t}e_{l}^{u}$, where $r+s+t+u=4$. The full expression is very complicated and is not given. Per Lemma \ref{lem:efro:hink} and \ref{lem:e4} and by conditional independence each of the conditional moments exist and are function of $l_{\boldsymbol{a_{i'}}}^{(\cdot k)}$, $i'=i,j,k,l$ and $k=2,3,4$. Which by Lemma \ref{lem:LikLim} $l_{\boldsymbol{a_{i'}}}^{(\cdot k)} = w_{i}\mu_{k} + o_{p}(1)$. The conclusion of this discussion is that $E[G(\boldsymbol{e})|\mathcal{A}]=G(\boldsymbol{w},\boldsymbol{\mu})[1+o_{p}(1)]$, where $\boldsymbol{\mu} = (\mu,\mu_{3},\mu_{4})^{T}$. Now combining all the terms we can write the conditional MSE as
\begin{align}
    {\rm{MSE}}[\hat{\boldsymbol{\beta}}|\mathcal{A}] &= J_{\mathcal{A}}^{-1} + J_{\mathcal{A}}^{-1}F^{T}\left\{\left[\frac{5}{4}I_{\mathcal{A}}^{-2}K_{\mathcal{A}}^{2} + \frac{1}{2}I_{\mathcal{A}}^{-1}L_{\mathcal{A}} \right] + G(\boldsymbol{w},\boldsymbol{\mu} ) \right\} F J_{\mathcal{A}}^{-1} + o_{p}(n^{-2}) \\
    &= J_{\mathcal{A}}^{-1} + (n\mu)^{-2} M^{-1} F^{T}\left\{ \left[\frac{5}{4}\frac{\mu_{3}^{2}}{\mu^{2}} + \frac{1}{2}\frac{\mu_{4}}{\mu}\right] + G(\boldsymbol{w},\boldsymbol{\mu} )  \right\}M^{-1} F + o_{p}(n^{-2}) \\
    &= J_{\mathcal{A}}^{-1} + n^{-2}G'(\boldsymbol{w},\boldsymbol{\mu}) + o_{p}(n^{-2}),
\end{align}
where $G'(\boldsymbol{w},\boldsymbol{\mu})$ is a finite matrix. $\square$

\end{supplement}

%%  MR numbers will be added by VTeX.                      %%
%%                                                         %%
%%  Use \cite{...} to cite references in text.             %%
%%                                                         %%
%%%%%%%%%%%%%%%%%%%%%%%%%%%%%%%%%%%%%%%%%%%%%%%%%%%%%%%%%%%%%

%% if your bibliography is in bibtex format, uncomment commands:
%\bibliographystyle{imsart-number} % Style BST file (imsart-number.bst or imsart-nameyear.bst)
%\bibliography{bibliography}       % Bibliography file (usually '*.bib')

%% or include bibliography directly:
\bibliographystyle{imsart-number}
\bibliography{bibtex_entries}

\end{document}